\documentclass[journal, 10pt]{IEEEtran}
\usepackage{algorithm,multirow}
\usepackage[noend]{algorithmic}
\usepackage{graphicx}
\usepackage{comment}
\usepackage{cite}
\usepackage{multirow}
\usepackage{amssymb,amsthm,mathrsfs}
\usepackage{amsfonts,epsf,graphicx,times,amsmath,multirow,cite}
\usepackage{float}
\usepackage{flexisym}
\usepackage[capitalise]{cleveref}
\usepackage{float}
\usepackage{verbatim}

\usepackage{amsthm}
\usepackage{subcaption}
\usepackage{color}

\usepackage{footnote}
\usepackage{footmisc}
\makesavenoteenv{tabular}
\makesavenoteenv{table}
\renewcommand{\thefootnote}{\alph{footnote}}
\newcommand{\astfootnote}[1]{
\let\oldthefootnote=\thefootnote
\setcounter{footnote}{0}
\renewcommand{\thefootnote}{\fnsymbol{footnote}}
\footnote{#1}
\let\thefootnote=\oldthefootnote
}

\usepackage{wrapfig}

\usepackage{url}

\graphicspath{{figs/}}
	\DeclareGraphicsExtensions{.pdf}

\usepackage{array}
\newcolumntype{L}[1]{>{\raggedright\let\newline\\\arraybackslash\hspace{0pt}}m{#1}}
\newcolumntype{C}[1]{>{\centering\let\newline\\\arraybackslash\hspace{0pt}}m{#1}}
\newcolumntype{R}[1]{>{\raggedleft\let\newline\\\arraybackslash\hspace{0pt}}m{#1}}

\makeatletter
\def\ps@headings{%
\def\@oddhead{\mbox{}\scriptsize\rightmark \hfil \thepage}%
\def\@evenhead{\scriptsize\thepage \hfil \leftmark\mbox{}}%
\def\@oddfoot{}%
\def\@evenfoot{}}
\makeatother
\usepackage [english]{babel}
\usepackage [autostyle, english = american]{csquotes}
\MakeOuterQuote{"}
\usepackage[final]{changes}
\definechangesauthor[name={Jose Terrero}, color=red]{JT}
\definechangesauthor[name={Yuqi Fu}, color=green]{yuqi}

\usepackage{caption}
\usepackage{subcaption}

\usepackage{array}
\usepackage{makecell}

\begin{document}

\date{}


\title{\huge Data-driven Analytical Models of COVID-2019 for Epidemic Prediction, Clinical Diagnosis, Policy Effectiveness and Contact Tracing: A Survey}

\author{

Ying Mao*\thanks{*Ying Mao is the corresponding author.},~
Susiyan Jiang,~
Daniel Nametz^\thanks{^Daniel Nametz is the lead research assistant.},~ \\
Yuxin Lin, Jake Hack, John Hensley, Ryan Monaghan, Tess Gutenbrunner 

\IEEEcompsocitemizethanks{
\IEEEcompsocthanksitem S. Jiang is with the Department of Pediatrics, New York Medical College, United States. E-mail: sjiang2@nymc.edu
\IEEEcompsocthanksitem Y. Mao, D. Nametz, J. Hack, J. Hensley, R. Monaghan, and T. Gutenbrunner are with the Department of Computer and Information Science at Fordham University in the New York City. E-mail: \{ymao41, dnametz, jhack, jhensley1, rmonaghan4, and tgutenbrunner\}@fordham.edu
\IEEEcompsocthanksitem Y. Lin is with the Community College Research Center, Columbia University, United States. E-mail: yl2879@tc.columbia.edu
}

}

\maketitle

\begin{abstract}

The widely spread CoronaVirus Disease (COVID)-19 is one of the worst infectious disease outbreaks in history and has become an emergency of primary international concern. 
As the pandemic evolves, academic communities have been actively involved in various capacities, including accurate epidemic
estimation,  fast clinical diagnosis, policy effectiveness evaluation and development of contract tracing technologies.
There are more than 23,000 academic papers on the COVID-19 outbreak, and this number is doubling every 20 days while the pandemic is still on-going~\cite{sciencemag}.
The literature, however, at its early stage, lacks a comprehensive survey from a data analytics perspective.
In this paper, we review the latest models for analyzing COVID-19 related data, conduct post-publication model evaluations and
cross-model comparisons, and collect data sources from different projects.

\end{abstract}


\section{Introduction}
\label{intro}

With over 7,800,000 cases and 430,000 deaths globally~\cite{jhu}, CoronaVirus Disease (COVID)-19, the disease caused by Severe Acute Respiratory Syndrome CoronaVirus (SARS-CoV)-2, is one of worst infectious disease outbreaks in history and has become an emergency of primary international concern. In mid-December 2019, the first COVID-19 case was detected in Wuhan, China, where it rapidly spread across the country and caused a pneumonia epidemic in early January 2020. The virus currently has spread to 140 other countries, including Japan, Italy, Brazil, and the USA, after infecting and causing the death of thousands of patients in China, with the number of confirmed new cases and deaths increasing every day~\cite{peng2020management, zhai2020epidemiology, pascarella2020covid}.

Hospitals and healthcare systems worldwide are under high stress and have already stepped up in unprecedented ways to face the challenges of COVID-19. For example, the first confirmed case of COVID-19 in the United States was reported in Snohomish County, Washington State~\cite{holshue2020first}. The genomic and epidemiological analyses of sequenced virus RNA recovered that in February 2020, community transmission of COVID-19 was detected in the western Washington region. Confirmed cases in the U.S. increased to 1,000 by March 11, to 100K by March 27, over 1 Million on April 28, and reaching 2 Million at the end of May~\cite{cdc}. In order to save lives and minimize the virus spread, hospitals have accelerated testing efforts and are treating hundreds of thousands of people worldwide.

The virus has influenced people’s daily life. To mitigate the spread of the disease, Wuhan in China Hubei province was placed under a strict lockdown on January 23 and reopened gradually after more than ten weeks. In the United States, California Gov. Gavin Newsom issued a stay-at-home order on March 19, and every state in the USA had restrictions in place by early April. The virus has also effectively grounded global economies to a halt. The U.S. unemployment rate had shot up from 3.8\% in February to 13.3\% in May~\cite{unemployment}, and the COVID-19 recession is predicted to be comparable to the Great Depression of the 1930s, where the unemployment rate was estimated to reach 25\%~\cite{recession}. 

To combat this ongoing crisis, many efforts have been made in developing accurate epidemic predictions, fast diagnosis solutions, effective policy implementations and efficient tracing systems. These projects, ranging from using different kinds of clinical data (chest CT image, X-Ray, laboratory findings, etc.) to generate fast screening methods, risk profiling, patient surveillance and tracking, and genetic network analysis, provide a snapshot of pandemic origins and social analytics. 
Many institutions have already developed COVID-19 tracking projects and generated data dashboards to help policymakers and the public understand the trajectory of the pandemic, compare each country or state's interventions and testing levels with case counts and death overtime, and make decisions for the path forward. 

While various datasets and analyses have been published by hospitals, institutions, governments and organizations globally, it lacks a comprehensive literature review and data collection from the analytical perspective to address the fragmented data
at the early stage of COVID-19 related researches. For example, a prediction model  published in early March showed
accurate estimation of infection numbers might not work well due to the fast evolution of the virus and government responses.
This manuscript demonstrates the latest datasets, prediction models, evaluations of policies, and tracing technologies for combating the challenges caused by COVID-19. The main contributions of this paper are summarized below.

\begin{itemize}

\item We provide an overview of data-driven COVID-19 studies from the perspectives of
epidemic prediction, clinical diagnosis, policy effectiveness, and contact tracing.

\item We conduct model studies with the latest data to evaluate
how good they perform since their publication date.

\item We collect the data sources and timeline of the key policies and combine them with multiple data sources to 
estimate the effectiveness.

\end{itemize}

The rest of this paper is organized as follows. In Section~\ref{infection},
we review the state-of-the-art models for epidemic prediction. 
In Section~\ref{clinical}, we report the analytical studies about clinical characteristics and diagnosis. 
We present the policy effectiveness researches in Section~\ref{policy}.
The latest technologies for COVID-19 related contact tracing are reviewed in Section~\ref{contact}.
Final, Section~\ref{con} concludes this.


\section{Epidemic Prediction}
\label{infection}

\begin{figure*}[!t]
\centering
         \begin{subfigure}[t]{0.24\linewidth}
\centering
         \includegraphics[width=\linewidth]{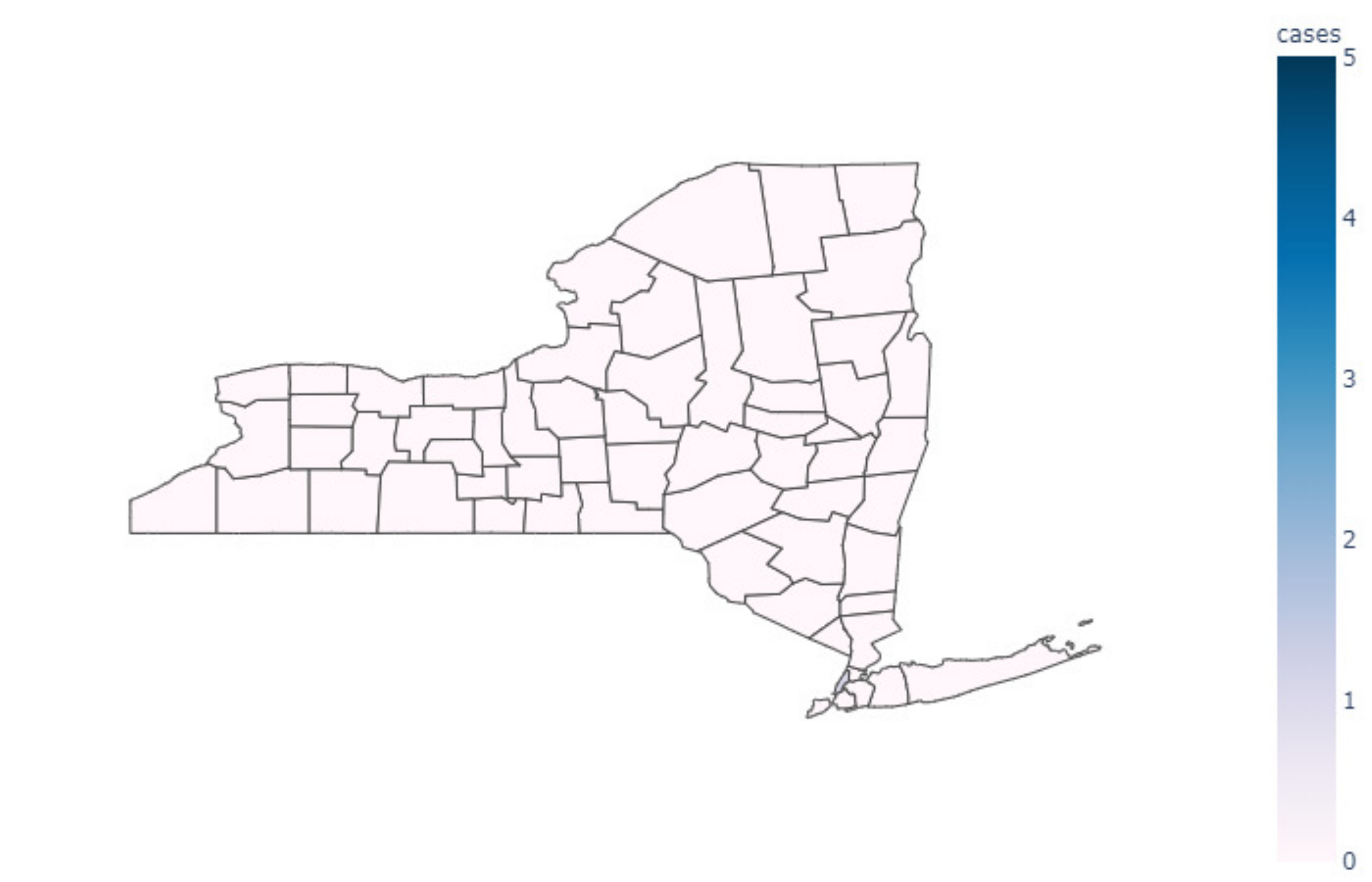}
\caption{March 1st, 2020}
      \label{3-01}
      \end{subfigure} 
      ~
      \begin{subfigure}[t]{0.24\linewidth}
\centering
      \includegraphics[width=\linewidth]{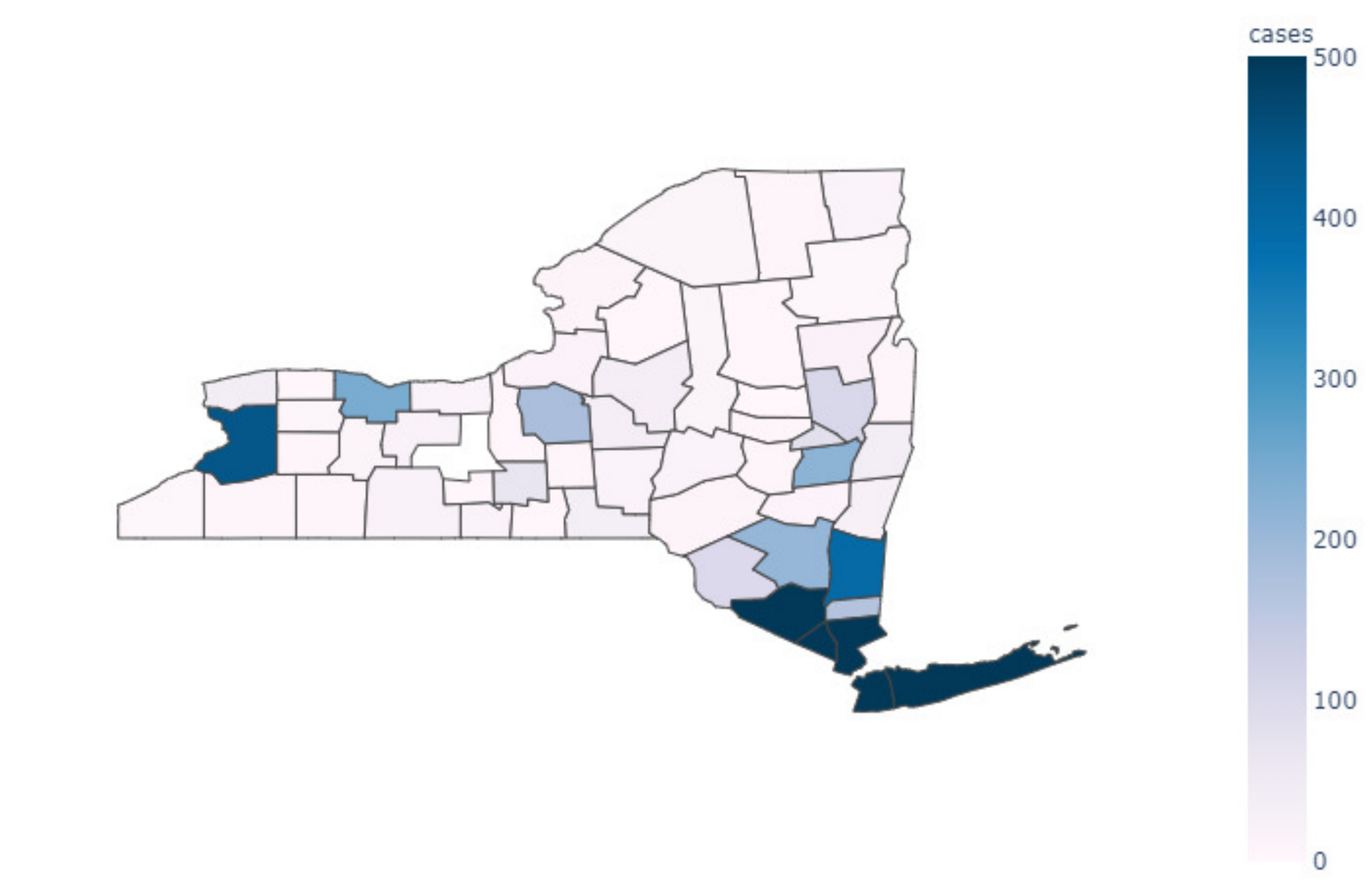}
\caption{March 30th, 2020}
      \label{3-30}
      \end{subfigure} %
      ~
      \begin{subfigure}[t]{0.24\linewidth}
\centering
      \includegraphics[width=\linewidth]{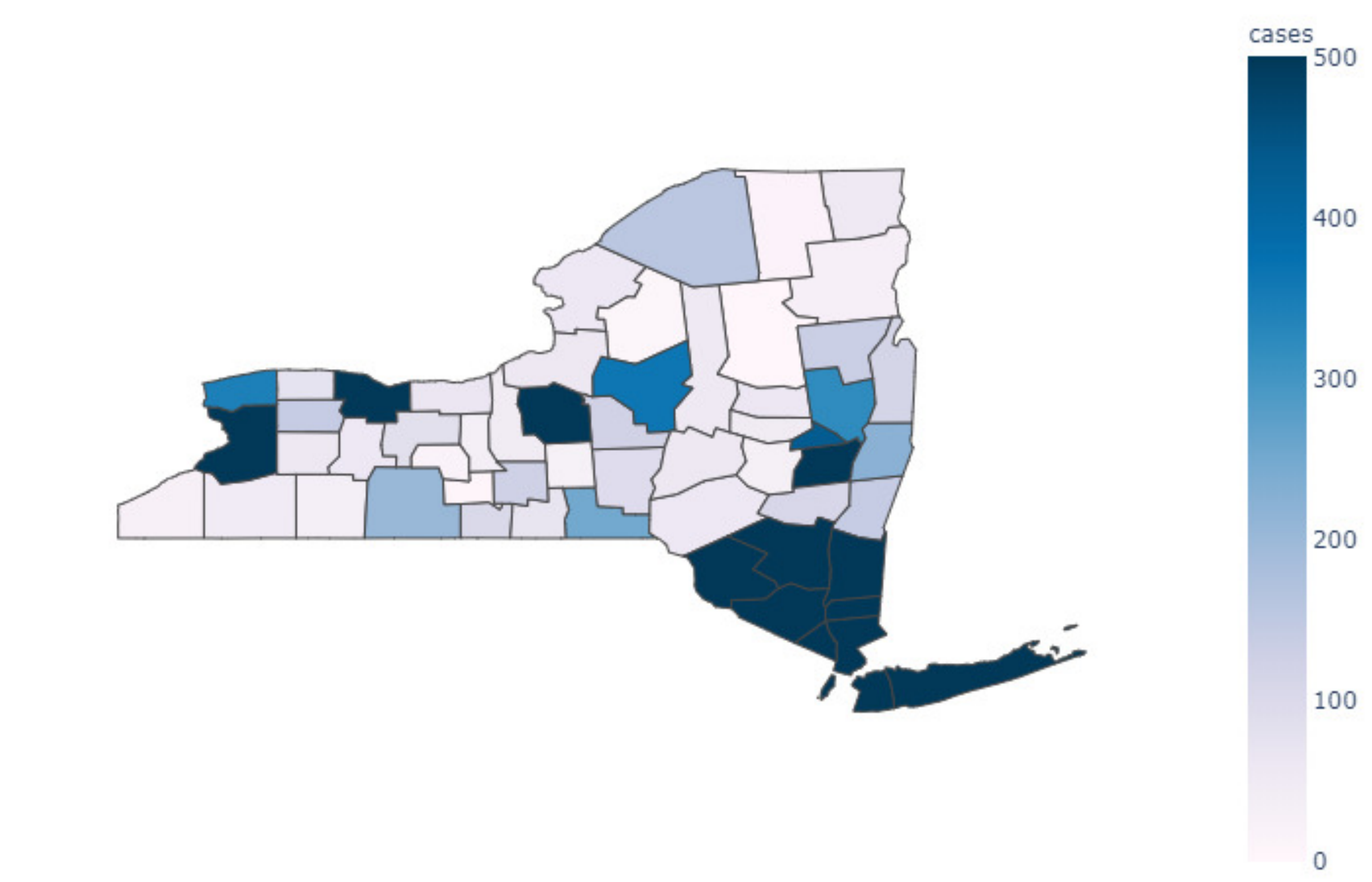}
\caption{April 25th, 2020}
      \label{4-25}
      \end{subfigure} %
      ~
      \begin{subfigure}[t]{0.24\linewidth}
\centering
      \includegraphics[width=\linewidth]{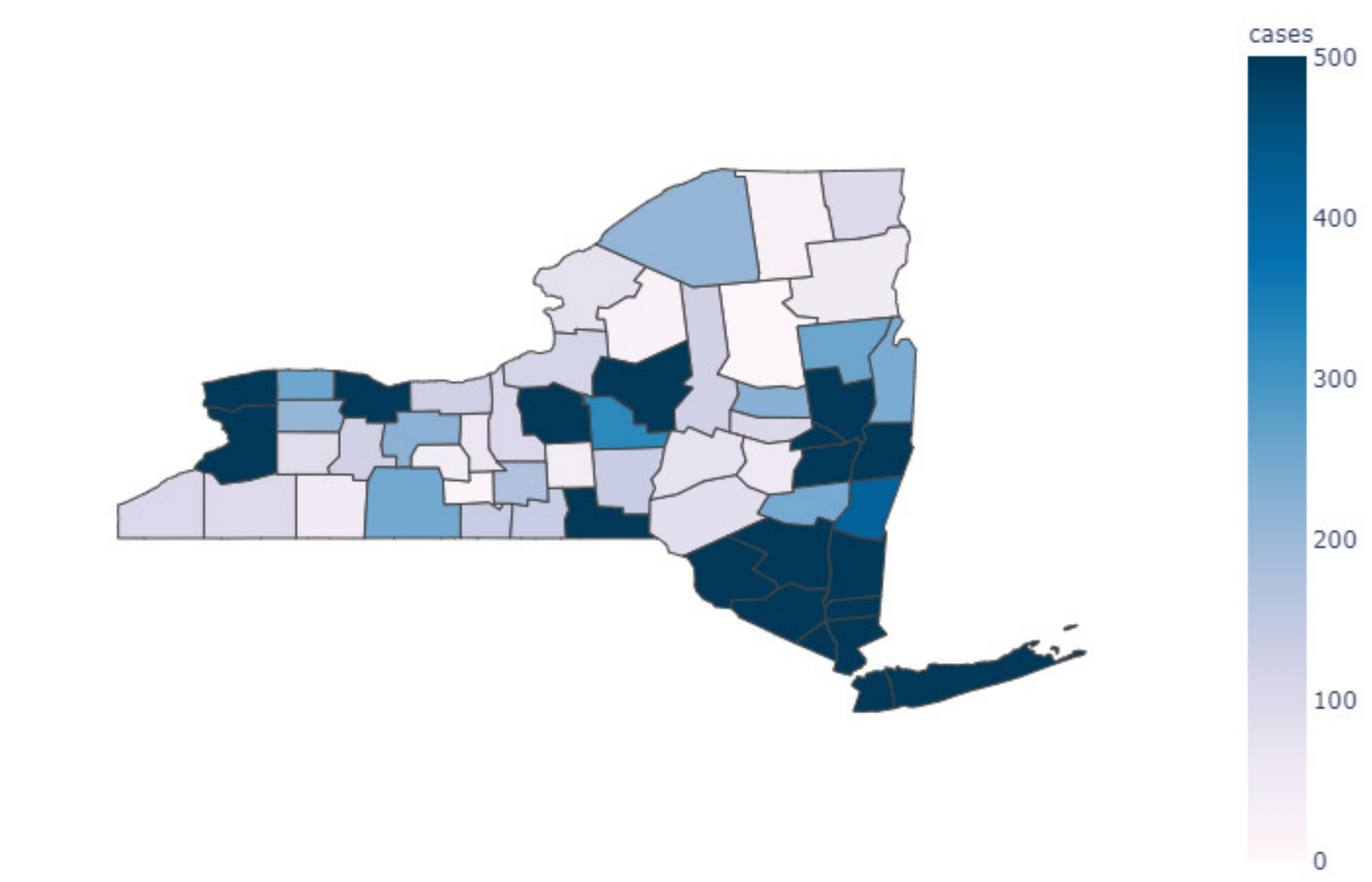}
\caption{June 7th, 2020}
      \label{6-07}
      \end{subfigure} %
\caption{Per-County Infection Map of New York State}  
\label{nystate}    
\end{figure*}

The tremendous increase in the number of infected patients with COVID-19 has
drained the healthcare systems globally. 
Based on the New York Times Data Set~\cite{nyt}, Fig.~\ref{nystate} illustrates how rapidly the virus spreads in the counties of New York State (NYS), which is the epicenter of the coronavirus in the United States. 
The heatmap figure plots the number of infections per each county in NYS.
On March 1st, there was only one confirmed COVID-19 case in NY state; however, the number increased to 
67462, 286901, 382879 on March 30, April 25 and June 7.  
An urgent need exists to accurately predict the epidemic.
Many efforts have been made to estimate the scale and time course of
epidemics, evaluating the effectiveness of public health interventions, and informing public health policies. 


\subsection{Prediction Models}

\subsubsection{Exponential Model}

Without effective responses (e.g. the early stages of a pandemic), the number of infected patients will grow exponentially
over the time. Given the initial time series data of diagnosed infections, we can get,

\begin{equation}
 I(t) = I(0) \times e^{rt}
\end{equation}
,where $I(t)$ is the number of diagnosed infections over the time and $r$ is the growth rate, which can be obtained
though observed data at the moment when executing the model.

The authors in~\cite{cassaro2020can, ranjan2020predictions} studied the exponential model, however, in practice, 
the prediction fails to deliver reliable numbers due to active responses from the government. 

\subsubsection{Logistic Model }

Unlike the exponential model that only works for the uncontrolled prevalence, the logistic growth model is approximately exponential at first, but the growth rate reduces as it approaches the model's upper bound, called the carrying capacity.
In the logistic model, the growth is given by~\cite{jia2020prediction},

\begin{equation}
I(t) = \frac{N}{1+e^{b-c(t-t_0)}} 
\end{equation}
, where $I(t)$ is the cumulative number of confirmed cases, $N$ is the predicted maximum number of confirmed cases (carrying capacity of the population),
$b$ and $c$ are fitting coefficients which can be obtained by using the existing data set, $t_0$ is the time when the first infection is observed and $t$ is the number of days since the first case.

Similar logistic growth and regression based models were developed to predict trends of the pandemic~\cite{zhou2020forecasting, wu2020generalized, tatrai2020covid, kriston2020projection, huang2020spatial, zhang2020predicting}. 
For example, authors~\cite{zhang2020predicting} proposed a segment Poisson model that coupled a power law with an exponential law
to estimate outbreaks. 
However, according to the latest evolution of COVID-19 worldwide, the model consistently under predicts the final epidemic size.

\subsubsection{SIR Model}

The Susceptible-Infectious-Recovered (SIR) is a compartmental model that describes the transmission of an infectious disease through individuals who pass through the following five states: susceptible, infectious, and recovered.
Their distributions can be given as follows~\cite{ranjan2020predictions},

\begin{equation}
\frac{dS(t)}{d(t)} = - \frac{\beta}{N}\times S \times I
\end{equation}
\begin{equation}
\frac{dI(t)}{d(t)} = (\frac{\beta}{N}\times S - \gamma)\times I
\end{equation}
\begin{equation}
\frac{dR(t)}{d(t)} = \gamma \times I
\end{equation}
, where $\beta$ is the transmission rate, $\gamma$ is the recover rate recovery, and $N = S+I+R$ is a constant
The basic reproduction number in SIR model is, 
\begin{equation}
R_0 = \frac{\beta}{\gamma} (1 - \frac{I_0}{N})
\end{equation}

As a popular base model for predicting COVID-19, SIR has many variations in the literature. For example, a modified Susceptible-Exposed-Infectious-Removed (SEIR) epidemiological model was proposed in~\cite{yang2020modified}, which introduced move-in, In(t) and move-out, Out(t) parameters to respect the mass population in Wuhan during the Chinese New Year. Additionally, a Stochastic SIR model (SSIR) was proposed in~\cite{simha2020simple} that takes the randomness into the prediction.

\subsubsection{MetaWards}

The author in~\cite{danon2009role} adapted an existing stochastic metapopulation model of disease transmission to predict the likely timing of the peak of the COVID-19 epidemic in England and Wales.
The population was divided into electoral wards in this model, and the author assumed that the individuals would contribute to the force of infection in their "home" ward during the night and their “work” ward during the day. To estimate potential decreased transmission rate during the summer months, the author replaced the constant transmission rate with a time-varying transmission rate. The equation of the transmission rate is,

\begin{equation}
r = \beta \times (1 - \frac{m}{2} \times (1 - cos\frac{2\pi \times t}{365}))
\end{equation}
, where $m$ represents the magnitude of the seasonal difference in transmission and ranges from 0 (no seasonality) to 1 (maximum seasonality with no transmission at the peak of the summer). 


However, besides the seasonal factors, the MetaWards model fails to consider COVID-19 interventions, which results in 
significantly overestimated trends of infections. 

\subsubsection{SIDARTHE}

\begin{figure}[ht]
\centering
         \includegraphics[width=\linewidth]{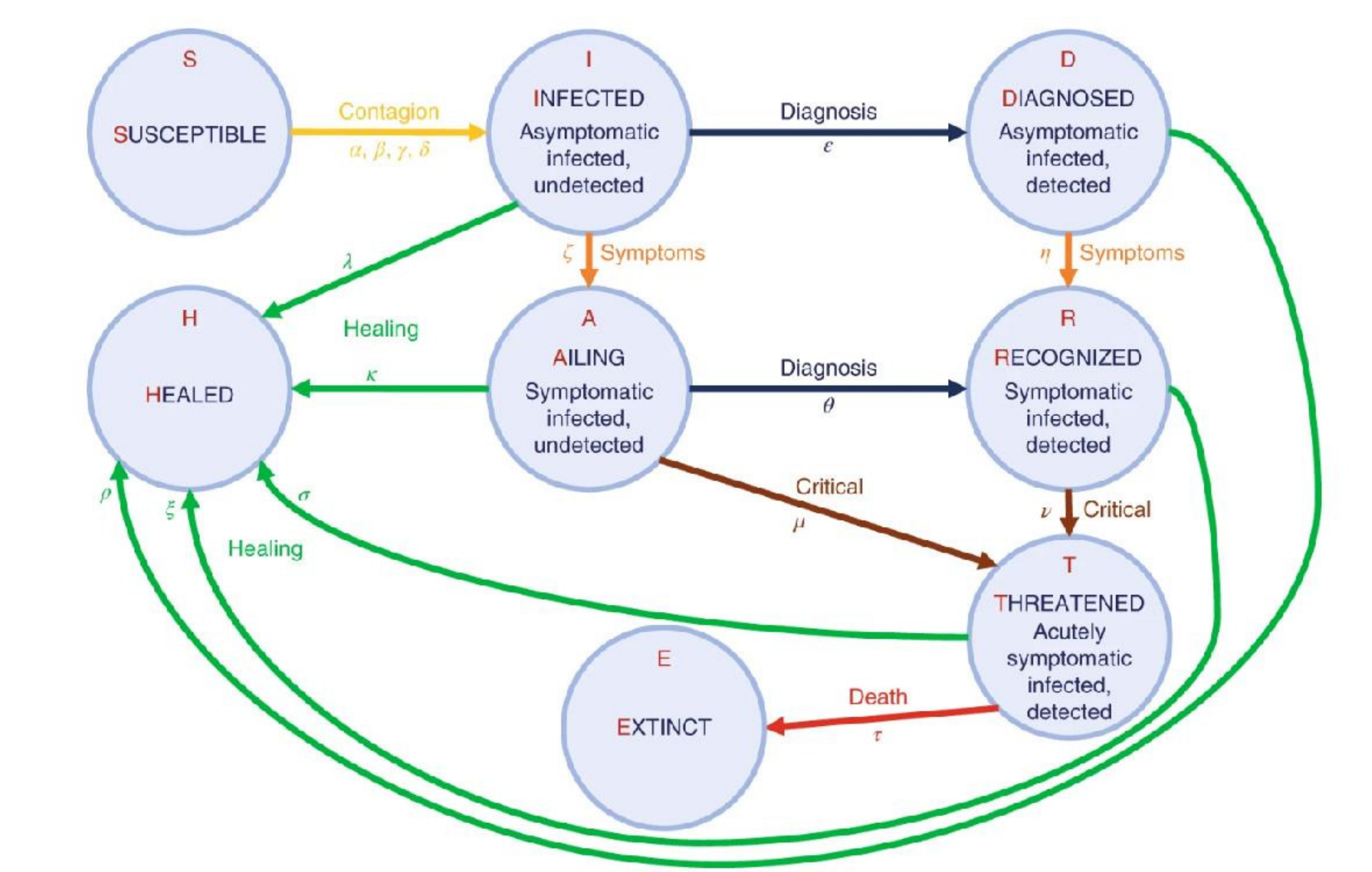}
\caption{SIDARTHE Model (Fig~.1 in ~\cite{giordano2020modelling})}
      \label{SIDARTHE-model}
\end{figure}

A more comprehensive model was proposed in~\cite{giordano2020modelling}. The model SIDARTHE considers multiple stages of the infection such that S, susceptible (uninfected); I, infected (asymptomatic or paucisymptomatic infected, undetected); D, diagnosed (asymptomatic infected, detected); A, ailing (symptomatic infected, undetected); R, recognized (symptomatic infected, detected); T, threatened (infected with life-threatening symptoms, detected); H, healed (recovered); E, extinct (dead).
Fig.~\ref{SIDARTHE-model} illustrates the stage transitions of SIDARTHE.

Specifically, it consists of eight ordinary differential equations, modeling the evolution of the population in each stage over time. 

\begin{equation}
\dot{S}(t) = -S(t) \times (\alpha I(t) + \beta D(t) + \gamma A(t) + \delta R(t))
\end{equation}

\begin{equation}
\begin{split}
\dot{I}(t) = S(t) \times (\alpha I(t) + \beta D(t) + \gamma A(t) \\ + \delta R(t))
 - (\epsilon + \zeta + \lambda)I(t)
\end{split}
\end{equation}

\begin{equation}
\dot{D}(t) = \epsilon \times I(t) - (\eta + \rho) D(t)
\end{equation}

\begin{equation}
\dot{A}(t) = \zeta \times I(t) - (\theta - \mu + \kappa) A(t)
\end{equation}

\begin{equation}
\dot{R}(t) = \eta D(t) + \theta A(t) - (\upsilon + \xi) R(t)
\end{equation}

\begin{equation}
\dot{T}(t)=\mu A(t)+v R(t)-(\sigma+\tau) T(t)
\end{equation}

\begin{equation}
\dot{H}(t)=\lambda I(t)+\rho D(t)+\kappa A(t)+\xi R(t)+\sigma T(t)
\end{equation}

\begin{equation}
\dot{E}(t)=\tau T(t)
\end{equation}
, where the state variables (upper Latin letters) are the population fraction of each stage and 
considered parameters (Greek letters) are positive numbers. The $\alpha, \beta, \gamma$ and $\delta$
are the transmission rate of contact between S and I, D, A and R. $\varepsilon \text { and } \theta$ 
are the detection probabilities of asymptomatic and symptomatic cases, respectively.
$\zeta$ and $\eta$ represent the probability rate at which an infected subject, respectively not aware and aware of being infected. 
$\mu$ and $v$ denote the probabilities of undetected and detected infected subjects that develop serious symptoms.
$\tau$ is the mortality rate. $\lambda,\kappa, \xi, \rho$ and $\sigma$ are the recovery rates for the patients in five classes (S, I, D, A and R).

\begin{figure*}[!t]
\centering
         \begin{subfigure}[t]{0.32\linewidth}
\centering
         \includegraphics[width=\linewidth]{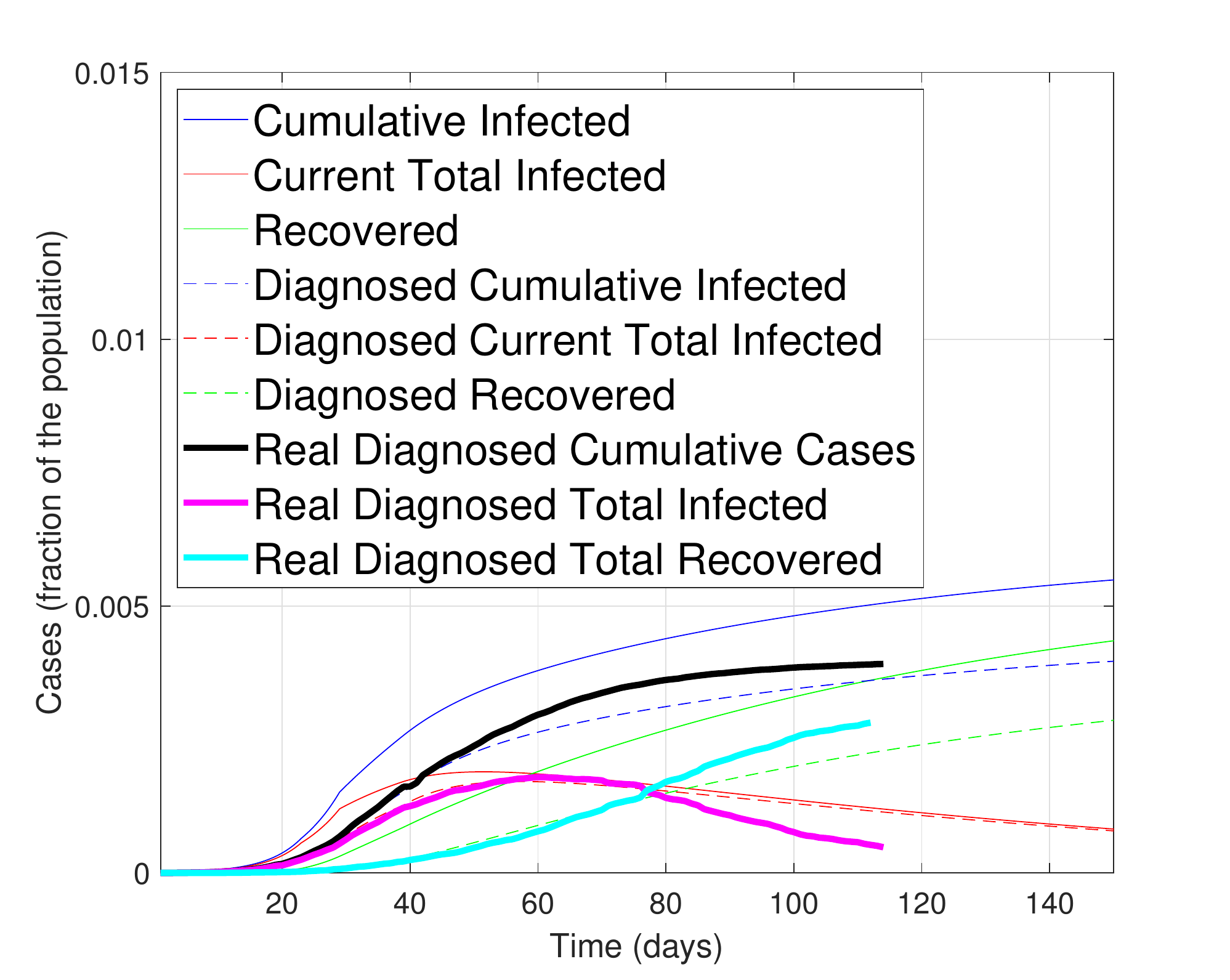}
\caption{Predicted epidemic evolution}
      \label{SIDARTHE-predict}
      \end{subfigure} 
      ~
      \begin{subfigure}[t]{0.32\linewidth}
\centering
      \includegraphics[width=\linewidth]{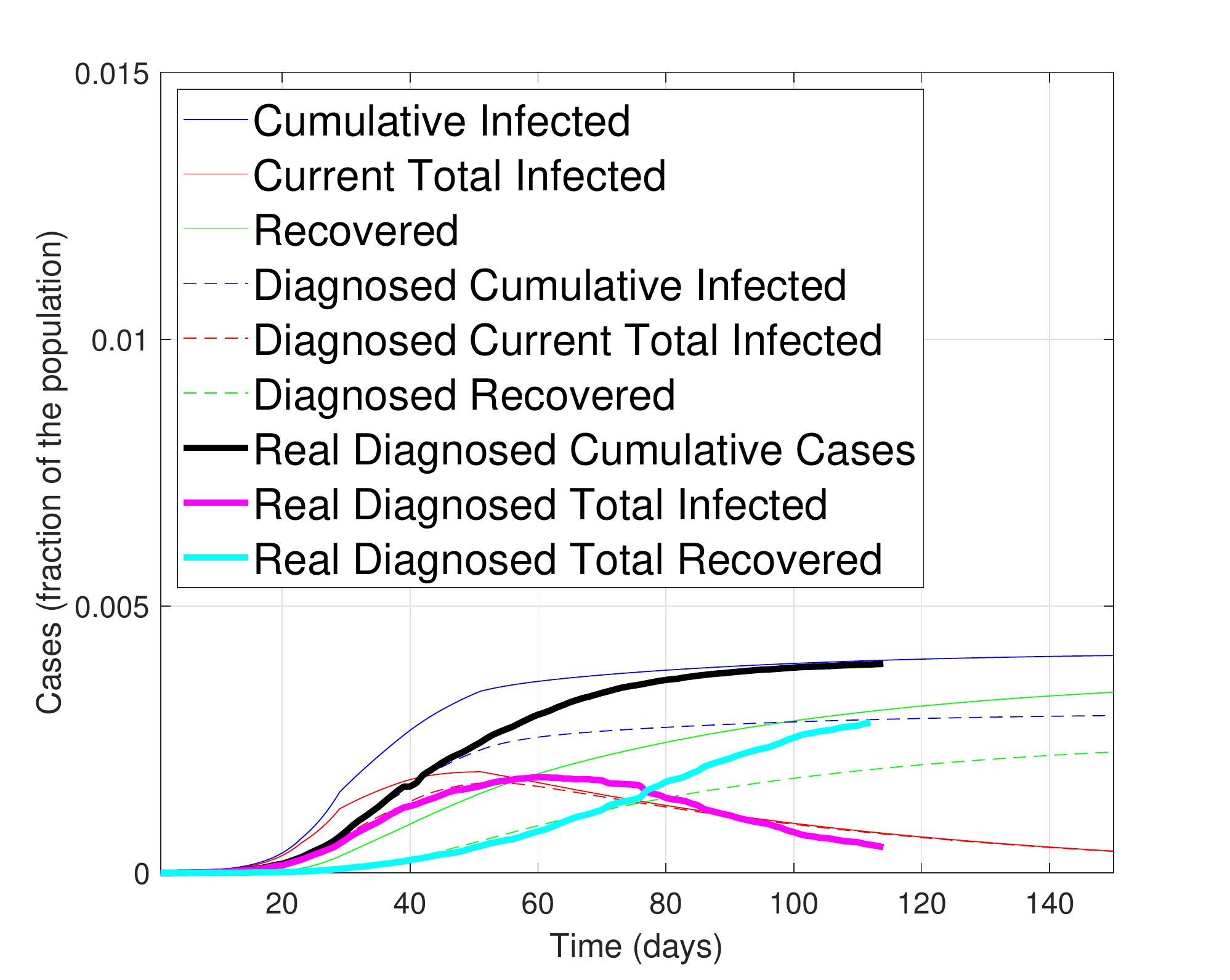}
\caption{The effect of lockdown}
      \label{SIDARTHE-effect}
      \end{subfigure} %
      ~
      \begin{subfigure}[t]{0.32\linewidth}
\centering
      \includegraphics[width=\linewidth]{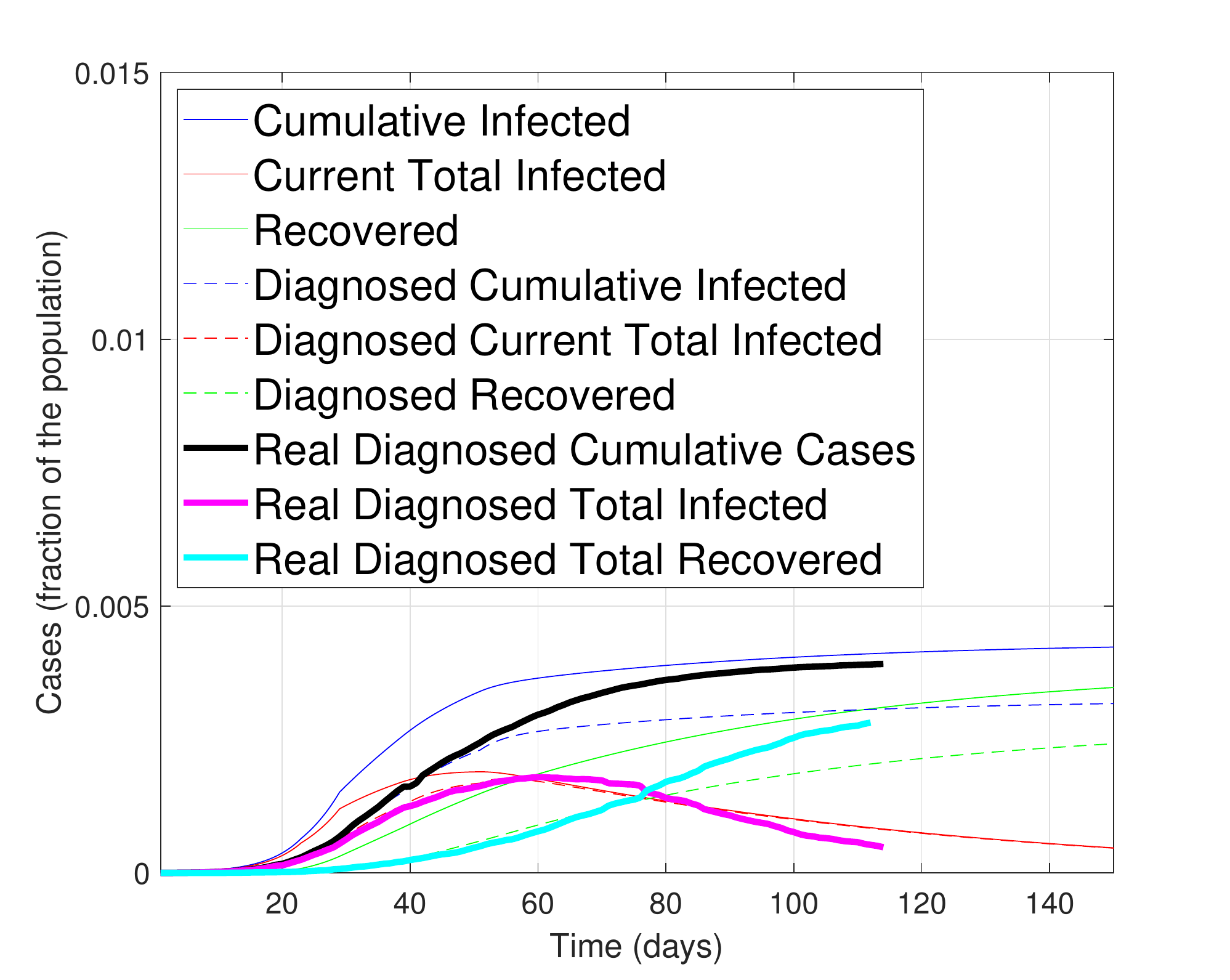}
\caption{The effect of testing}
      \label{SIDARTHE-test}
      \end{subfigure} %
\caption{Followup of the SIDARTHE model with 70 days post-publication data}  
\label{SIDARTHE}    
\end{figure*}

\subsection{Post Publication Model Evaluation}

The SIDARTHE model utilizes much more parameters than the previously mentioned models. 
Based on the code provided by authors, Fig.~\ref{SIDARTHE} plots the midterm evolution of the pandemic in Italy, which 
contains three additional curves with the latest data, Real Diagnosed Cumulative Cases (black), Real Diagnosed Total Infected (pink) and Real Diagnosed Total Recovered (light blue). Please note that the model uses data from 0 to 45 days to obtain the best parameter set. The solid lines (except the additional three)  are estimates of the actual pandemic, and the dotted lines are the estimates of the diagnosed pandemic.

As we can see from Fig.~\ref{SIDARTHE-predict}, the predicted values of diagnosed cumulative cases are always lower than the real values from 45 to 116. However, the difference between the two curves is reducing over time. It is likely that the restrictions were taking effect gradually. Fig.~\ref{SIDARTHE-effect} plots the prediction of a stricter lockdown enforcement.
The curve flattens quicker than the real case (black curve). Fig.~\ref{SIDARTHE-test} assumes a milder lockdown with widespread testing and contact tracing. It fits well with the real curve with a stable gap, which reflects the shortage of tests and contact tracing. Additionally, the real diagnosed total number of recovered cases are more than expected. This is due to the fact that more resources, such as ventilators, are available to the patients. 

Almost all reviewed models fit the curve very well before their publication date; however, the performance of their estimations is 
yet to be discovered. In Table~\ref{table:prediction}, we present the post -publication statistics of 13 models and 17 countries published from Feb. 28 to Jun. 6. For the papers, Table~\ref{table:prediction} collects the Predicted Peak Date (P. Peak Date), Predicted Peak Value (P. Peak Value, largest number daily new cases) and Predict Size (P. Size, e.g. the predicted final size of the pandemic). From the database of World Health Organization~\cite{who}, we queried the data (on Jun. 18, 2020) of Real Peak Date (R. Peak Date, the date of largest daily new cases), Real Peak Value (R. Peak Value, the largest number of daily new cases) and
Current Size (the current total infections). We can see from the table that models~\cite{jia2020prediction, yang2020modified}, which predict China will produce reasonably good results on the predicted peak date and size. This is because, when they published in late February and mid March, the curve of infections in China had already been flattened and was reduced, which means China was on track towards the end of the pandemic and thus, the models have enough data for training and fitting the curve. 
Except for China, COVID-19 was still quickly spreading in many countries in late March and early April. 
For example, on April 8, the model in~\cite{zhou2020forecasting} predicts that the final pandemic size would be
71950, 36240, 10420, 85750, 85750, 41850, 61420, 1560, for Italy, Iran, South Korea, Germany, France, USA, Spain and Japan, respectively. 
With the latest data on Jun. 18, however, most of these numbers are significantly underestimated, 230\%,
438\%, 17.6\%, 118\%, 313\%, 4900\%, 297\%, 1000\% for these counties. The most accurate prediction was for South Korea since, at the time of its publication, the trend in South Korea was clear enough for the model.

\begin{table*}[ht]
	\centering
	\caption{Epidemic Prediction Post Publication Evaluation (as of June 18 2020)}
	\scalebox{0.95}{
		
		\begin{tabular}{ | c | c | c | c | c | c | c | c | c |}				
			\hline
			Model & Publish Date  		&  Region & P. Peak Date & P. Peak Value & P. Size & R. Peak Date  & R. Peak Value & Current Size  \\ \hline	
			\makecell{SEIR~\cite{yang2020modified} \\ LSTM~\cite{yang2020modified}}  & 02/28/20  & China	 & \makecell{02/18/20 \\ 02/04/20} &  \makecell{4169 \\ 3886}	& \makecell{122122 \\ 95811}   & 02/13 & 15152 & 84867\\ \hline
            
            Logistic Model~\cite{jia2020prediction} & 03/16/20 & China & 02/06/20 & $\sim$8000 & 80261 & 02/13 & 15152 & 84867 \\ \hline
            
            SIR~\cite{roda2020difficult}  & 03/16/20 & China & 02/27/20 & N/A & 120000 & 02/13 & 15152 & 84867 \\ \hline
			
			SIDARTHE~\cite{giordano2020modelling} & 03/22/20 & Italy & 03/15 & N/A	& 181080 	& 03/21 & 6557 & 237500 \\ \hline
			
			SIR~\cite{ranjan2020predictions} & 04/06/20 & India & 04/12/20 & 1500 & 13000 & 06/14 & 11929 & 354065 \\ \hline
			
			Adjusted SEIR~\cite{zhou2020forecasting}  & 04/08/20 

& \makecell{
Italy\\
Iran \\
South Korea \\
Germany \\ 
France \\
USA \\
Spain \\
Japan
}

&  \makecell{
03/31 \\
03/31 \\
03/30 \\
04/03 \\
04/02 \\
04/07 \\
04/01 \\
03/30 
}

& N/A

& \makecell{
71950 \\ 
36240 \\ 
10420 \\
85750 \\ 
36980 \\ 
41850 \\ 
61420 \\ 
1560 }

& \makecell{
03/21 \\
06/04 \\
03/01 \\
03/28 \\
04/01 \\
04/26 \\
04/01 \\
04/12 }

& \makecell{
6557 \\
3574 \\
1062 \\
6294 \\
7500 \\ 
38509 \\
9222 \\
743 }

& \makecell{
237500 \\
195051 \\
12257 \\
187184 \\  
153045 \\
2098106 \\
244328 \\
17668 } \\ \hline

            \makecell{Segmented Poisson~\cite{zhang2020predicting}} & 04/20/20 & \makecell{France \\ Italy \\ USA \\ UK \\ Germany 
		    \\ Canada} & \makecell{04/07 \\ 03/26 \\ 04/07 \\ 04/09 \\ 03/31 \\ 04/06} & \makecell{$\sim$7500 \\ $\sim$5500 \\ $\sim$32500 \\ $\sim$4800 \\ $\sim$5500 \\ $\sim$1350} & \makecell{219583	\\ 172451 \\ 835158	\\ 133206 \\ 159437 \\ 33948	}
	    	&  \makecell{04/01 \\ 03/21 \\ 04/26 \\ 04/12 \\ 03/28 \\ 05/04}  &  \makecell{7500 \\ 6557 \\ 38509 \\ 8719 \\ 6294 \\ 3793}  & \makecell{153045 \\ 237500 \\ 2098106 \\ 298140 \\ 187184 \\ 99147} \\ \hline
	    	
	    	Gaussian\cite{barmparis2020estimating} & 04/27/20 & \makecell{
Greece \\
Netherlands \\
Germany \\
Italy \\
Spain \\ 
France \\
UK \\ 
USA 
}

& \makecell{
04/03 \\
03/31 \\
04/02 \\
03/26 \\
03/31 \\
04/05 \\
04/12 \\
04/05 
}

& \makecell{
$\sim$100 \\
$\sim$1100 \\
$\sim$6100 \\
$\sim$5950 \\
$\sim$8500 \\
$\sim$5500 \\
$\sim$6500 \\
$\sim$30000 }

& \makecell {
2811 \\
23713 \\
14003 \\
156975 \\ 
173525 \\ 
141973 \\ 
165443 \\
654207 
}

& \makecell {
04/21 \\
04/10 \\
03/28 \\
03/21 \\
04/01 \\
04/01 \\
04/12 \\
04/26
}

& \makecell {
156 \\
1335 \\
6294 \\ 
6557 \\
9222 \\
7500 \\
8719 \\
38509 
}

& \makecell {
3203 \\
49412 \\
187184 \\
237500 \\
244328 \\
153045 \\
298140 \\
2098106
} \\ \hline
            Modified SEIR~\cite{lopez2020modified} & 04/29/20 & \makecell{Spain \\ Italy} & \makecell{04/29 \\ 04/25} & N/A & 
		    \makecell{$\sim$ 125000 \\ $\sim$ 10000} & \makecell{04/01 \\ 03/21} & \makecell{9222 \\ 6557} & \makecell{244328 \\ 237500} \\ \hline
		
		    SEIRQRP~\cite{xu2020forecast} & 04/29/20 & USA & 05/18 & N/A & 820000 & 04/26 & 38509 & 2098106 \\ \hline

			Nonlinear LR model~\cite{bashir2020trend}	 & 05/13/20	& \makecell{Pakistan \\ USA \\ Italy \\ Spain}
			& \makecell{06/04 \\ 04/30 \\  04/28 \\ 05/04} 
			& \makecell{48000 \\ $\sim$ 1100000 \\ 19700 \\ 23600}
			& N/A & 	\makecell{06/13 \\ 04/26 \\ 03/21 \\ 04/01}
			& \makecell{6884 \\ 38509 \\ 6557 \\ 9222}
			& \makecell{154760 \\ 2098106 \\ 237500 \\ 244328} \\ \hline

			SIR~\cite{boudrioua2020predicting} & 06/06/20 & Algeria  &	 4/13	& 106 & 244400 & 04/02 & 263 & 11147 \\ \hline
			 
		\end{tabular}	
	
	}
	
	\label{table:prediction}
\end{table*}

\section{Clinical Characteristics and Diagnosis}
\label{clinical}

The purpose of this retrospective cohort study is to seek a faster and more reliable diagnosis method of COVID-19 and acquire more accurate conclusions concerning the clinical characteristics and mortality risk factors for patients with confirmed COVID-19 infection.
In this section, we review the literature from traditional meta analysis and artificial intelligence aided analysis.

\subsection{Meta Analysis} The authors in~\cite{sun2020clinical, parohan2020risk} screened medical databases from PubMed~\cite{pubmed}, Cochrane Library~\cite{cochranelibrary}, Embase databases~\cite{embase}, Scopus~\cite{scopus}, and Google scholar~\cite{googlescholar}. They collected the relevant literature dated up to February 24~~\cite{sun2020clinical} and May 1, 2020~\cite{parohan2020risk}, and then proposed a meta-analysis of a quantitative, formal procedure that aggregated, integrated, and reanalyzed the results of several independent studies. 

As a subset of the systematic review, meta-analyses attempt to collate empirical evidence fitting previously specified criteria to provide a more precise estimate of the effect of treatment or the risk factors concerning a disease~\cite{haidich2010meta}. The PRISMA (Preferred Reporting Items for Systematic reviews and Meta-analyses), which contains a 27-item checklist and a four-phase flow diagram,  is a guide to improve the reporting of systematic reviews and meta-analyses~\cite{liberati2009prisma}. The forest plot is commonly used to present the results of meta-analyses, where each study is shown with its effect size and corresponding 95\% confidence interval. The risk ratio (or relative risk) and the odds ratio are the two most common measures of effect used for dichotomous data in meta-analysis, while the standardized mean difference (SMD) estimation is the dominant method used for continuous data. The random-effects model in meta-analysis assumes the true treatment effect differs across studies and could generate an estimate of the average treatment effect~\cite{riley2011interpretation}. The greatest benefit of meta-analysis is the ability to examine the degree of heterogeneity among studies. A statistical test such as Cochran’s $\mathcal{X}^2$test or the Q-test is used to indicate the extent of heterogeneity. The author in~\cite{higgins2002quantifying} developed measures for the impact of heterogeneity and proposed three suitable statistics: 
\begin{equation}
H^{2}=\frac{Q}{k-1}
\end{equation}
, where Q is $\mathcal{X}^2$ heterogeneity statistic, and $k-1$ is its degrees of freedom;
\begin{equation}
R^{2}=\frac{v_{R}}{v_{F}}
\end{equation}
, which represents the ratio of the standard error of the underlying mean from a random effects meta-analysis to the standard error of a fixed effect meta-analytic estimate;

\begin{equation}
I^{2}=\frac{H^{2}-1}{H^{2}}
\end{equation}
, the inconsistency index that describes the proportion of total variation in study estimates due to heterogeneity.

The Newcastle-Ottawa Scale (NOS)~\cite{stang2010critical} was used to evaluate all literature, with the highest quality of literature scoring nine stars. Articles with the NOS score of higher than five stars were considered high-quality publications in the study. The random-effects model for meta-analysis was used to reduce the influence of heterogeneity between the included studies in the final conclusion~\cite{sun2020clinical}.

A total of 284 articles were retrieved, where 39 papers were eliminated due to repeated retrieval, 212 papers after reading abstracts, and 23 after reading the full text, a total of 10 articles of literature~\cite{huang2020clinical, wang2020clinical, chen2020epidemiological, guan2020clinical, chen2020analysis, sun2020early, yang2020epidemiological, li2020epidemiological, china2019novel, xu2020clinical}, including data from 50,466 patients were analyzed in author ~\cite{sun2020clinical}’s research. Original data were transformed by the double arcsine method to make them conform to the normal distribution, and the initial conclusion was then restored via the formula 
$P= (sin(\frac{tp}{2}))^2$ 
to reach the final conclusion. The Egger test with $P < .05$ was performed in response to publication bias, where the values larger than were considered as demonstrating no publication bias. The statistical software Stata version 12.0 was used to carry out the single-arm meta-analysis, and the results were presented in Table 1 (with Egger test results, which indicates there existed a publication bias in the meta-analysis of ARDS (Acute Respiratory Distress Syndrome) group ($P = 0.008$)).

\begin{table}[ht]
	\centering
	\caption{Clinical Characteristics for Patients with Confirmed COVID-19 via Meta-analysis~\cite{sun2020clinical}}
	\scalebox{0.95}{
		
		\begin{tabular}{ | c | c | c | c | }				
			\hline
			Symptom & Meta-analysis & Adjusted results & P  \\ \hline	
			Fever & \makecell{2.47 \\ 95\% CI: 2.26 - 2.67} & \makecell{0.891 \\ 95\% CI: 0.818 - 0.945} & 0.866 \\ \hline
			Cough & \makecell{2.03 \\ 95\% CI: 1.89 - 2.17} & \makecell{0.722 \\ 95\% CI: 0.657 - 0.782} & 0.278 \\ \hline
			\makecell {Muscle soreness \\ or fatigue} & \makecell{1.42 \\ 95\% CI: 0.96 - 1.88} & \makecell{0.425 \\ 95\% CI: 0.213 - 0.652} & 0.09 \\ \hline
			ARDS & \makecell{0.79 \\ 95\% CI: 0.43 - 1.15} & \makecell{0.148 \\ 95\% CI: 0.046 - 0.296} & 0.008 \\ \hline
			\makecell{Abnormal \\ chest CT} & \makecell{2.77 \\ 95\% CI: 2.57 - 2.97} & \makecell{0.966, \\ 95\% CI: 0.921 - 0.993} & 0.908 \\ \hline
			\makecell{Patient in \\ critical condition} & \makecell{0.88 \\ 95\% CI: 0.73 - 1.03} & \makecell{0.181 \\ 95\% CI:0.127 - 0.243} & 0.826 \\ \hline
			Death of patient & \makecell{0.42 \\ 95\% CI: 0.33 - 0.50} & \makecell{0.043 \\ 95\% CI: 0.027 - 0.061} & 0.258 \\ \hline
		\end{tabular}	
	}
	
	\label{table:metaanalysis}
\end{table}

\begin{figure}[ht]
\centering
         \includegraphics[width=\linewidth]{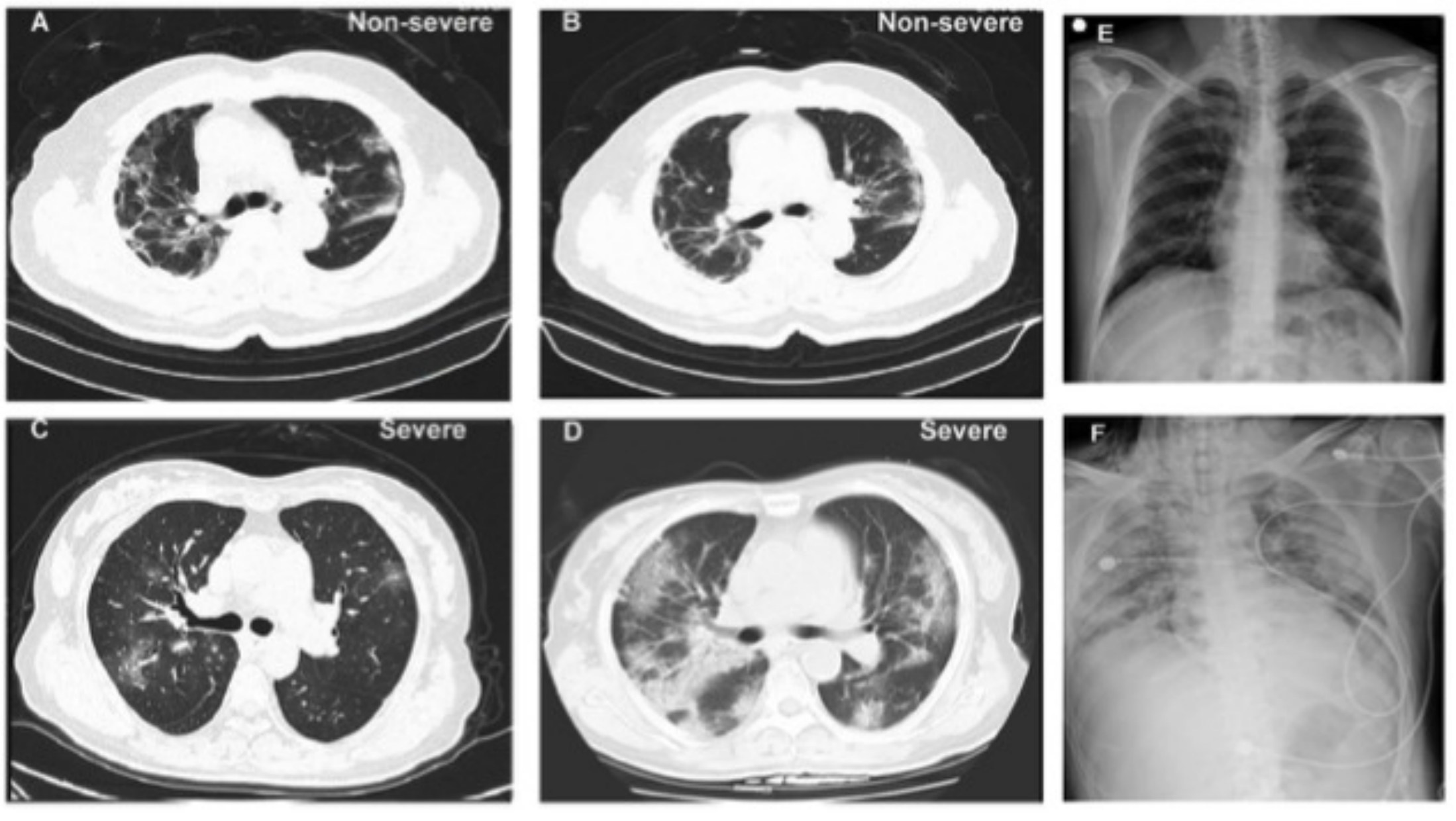}
\caption{Representative chest radiographic manifestations in a non-severe and a severe case with COVID-19~\cite{guan2020clinical}}
      \label{chestct}
\end{figure}

\begin{figure}[ht]
\centering
         \includegraphics[width=\linewidth]{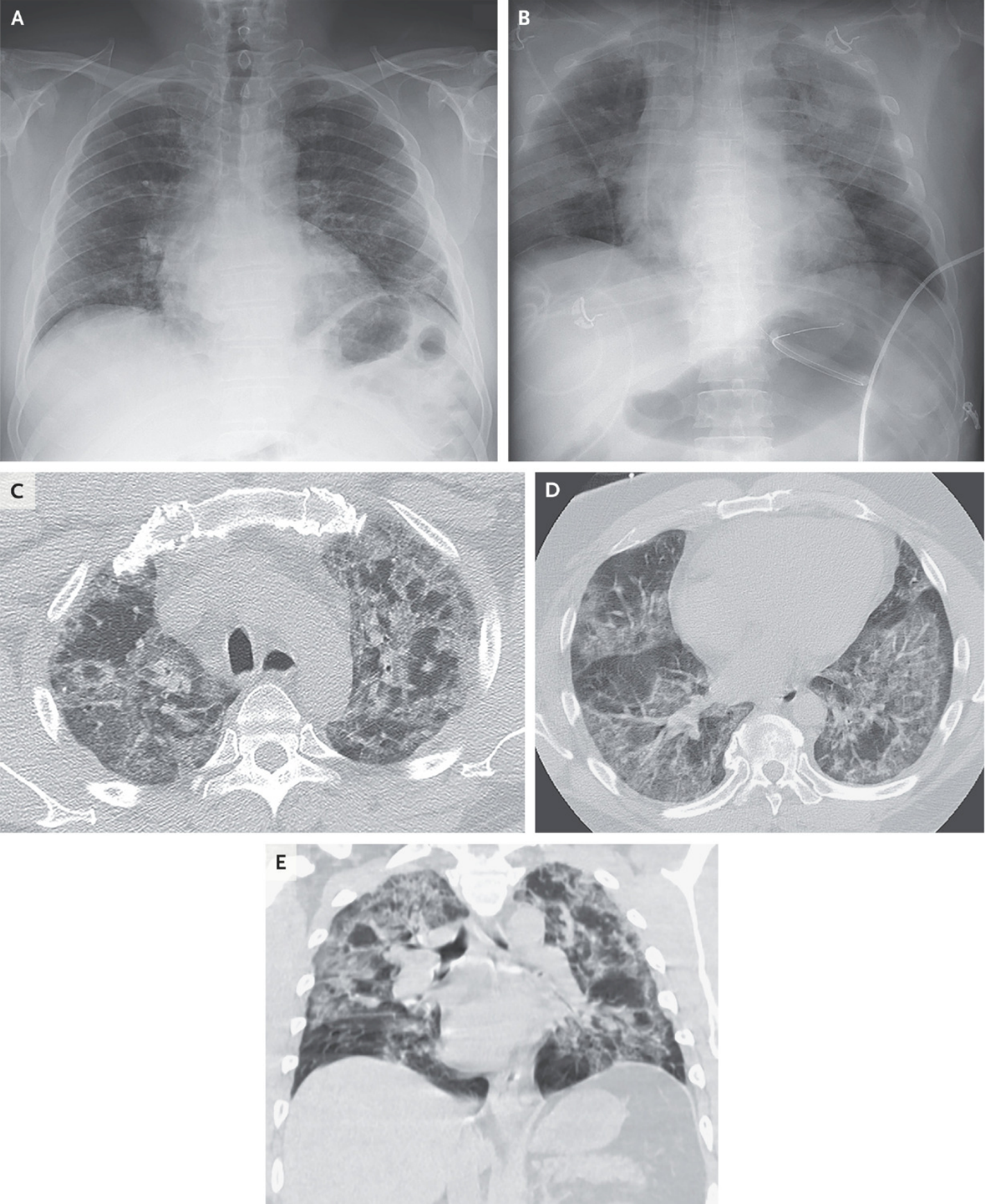}
\caption{Representative Chest Radiographs and CT Images of a Critically Ill COVID-19 patient in Seattle Region USA~\cite{bhatraju2020covid}}
      \label{chestct2}
\end{figure}

The three most common symptoms among people who were hospitalized with confirmed COVID-19 infection are fever (89.1\%), cough (72.2\%), and muscle or general fatigue (42.5\%). Diarrhea, hemoptysis, headache, sore throat, shock, and other symptoms are rare. 14.8\% of patients had ARDS; 18.1\% of all infected cases were defined as severe cases, and the mortality rate was 4.3\%. Chest CT scans were generally performed at the time of admission, and almost all patients (96.6\%) revealed abnormal results. The representative radiology findings in COVID-19 patients are shown in Fig.~\ref{chestct} and Figure 2~\ref{chestct2}. The common pattern on chest CT scans for patients with COVID-19 infection were ground-glass opacity and bilateral patchy shadowing~Fig.~\ref{chestct}, and the bilateral multiple lobular and subsegmental areas of consolidation were found on the typical chest CT images of severe cases~\cite{huang2020clinical}.

The authors in~\cite{parohan2020risk} systematically reviewed the present evidences towards the association between age, gender, hypertension, diabetes, chronic obstructive pulmonary disease (COPD), cardiovascular disease (CVD), and risk of death due to COVID-19 infection. They summarized the available findings by meta-analysis. The classic Cochran’s Q test~\cite{patil1975cochran} was performed to examine the heterogeneity across studies, where $I^2 \geq 50\%$ was considered to demonstrate such heterogeneity. The formal test of Egger was used, and all statistical analyses were conducted by Stata, version 14.0. 

A total of 14 studies (twelve conducted in China~\cite{guan2020clinical, zhou2020clinical, wu2020risk, caramelo2020estimation, cheng2020kidney, su2020risk, wang2020clinical, chen2020risk, du2020predictors, liu2020neutrophil, shi2020association, wang2020coronavirus}, one in Italy~\cite{colombi2020well}, and one in Iran~\cite{nikpouraghdam2020epidemiological}) with 29,909 COVID-19 infected patients and 1,445 cases of death were included in the research~\cite{parohan2020risk}, and the meta-analysis results are presented in Table~\ref{table:single}. The results of Egger’s test demonstrated that the hypothesize on the association of demographic characteristics and comorbidities with COVID-19 mortality did not depend on a single study. The authors~\cite{parohan2020risk} findings supported the hypothesis that patients who were of ages older than 65 years, male, with coexisting disorders including hypertension, CVDs, diabetes, COPD, and cancer were associated with higher risk of mortality from COVID-19 infection.

\begin{table}[ht]
	\centering
	\caption{Mortality Risk Factors for patients with confirmed COVID-19 via Meta-analysis~\cite{parohan2020risk}}
	\scalebox{0.95}{
		
		\begin{tabular}{ | c | c | c | c | c |}				
			\hline
			Study-ID & \makecell{pooled \\ Odds Ratios} & $I^2$(\%) &  \makecell{p \\ Hetero.} & \makecell{p \\ Egger} \\ \hline		\makecell{Age \\ $\geq$ 65 v.s. $<$65} & \makecell{4.59 \\ 95\% CI: 2.61 - 8.04} & 67.10 & 0.01 & 0.185 \\ \hline
			\makecell{Gender \\ Male v.s. Female} & \makecell{1.50 \\ 95\% CIs: 1.06 - 2.12} & 76.30 & 0.002 & 0.388 \\ \hline
			\makecell{Hypertension \\ Yes v.s. No} & \makecell{2.70 \\ 95\% CIs: 1.40 – 5.24} & 92.6 & $<$0.001 & 0.065 \\ \hline
			\makecell{Cardiovascular \\ diseases, Yes v.s. No} & \makecell{3.72 \\ 95\% CIs: 1.77 – 7.83} & 89.10 & $<$0.001 & 0.068 \\ \hline
			\makecell{Diabetes \\ Yes v.s. No} & \makecell{2.41 \\ 95\% CIs: 1.05 – 5.51} & 93.60 & $<$0.001 & 0.117 \\ \hline
			\makecell{Chronic obstructive \\  pulmonary disease \\ (COPD), Yes v.s. No} & \makecell{3.53 \\ 95\% CI: 1.79 – 6.96} & 72.20 & 0.001 & 0.178 \\ \hline
			\makecell{Cancer \\ Yes v.s. No} & \makecell{3.04 \\  95\% CIs: 1.80 – 5.14} & 41.60 & 0.114	& 0.054 \\ \hline

		\end{tabular}	
	}
	
	\label{table:single}
\end{table}

\subsection{Artificial Intelligence Aided Analysis}

A confirmed case of COVID-19 infection is routinely defined as a positive result on high-throughput sequencing or RT-PCR assay of nasal and pharyngeal swab specimens~\cite{guan2020clinical}. However, the RT-PCR test has three limitations: 
\begin{itemize}
\item The process is very slow and can take up to two days to complete.
\item The serial testing may be required to eliminate the possibility of false negative results.
\item In some areas, there exists a shortage of RT-PCR test kits
\end{itemize}
Those challenges underscore the urgent need for alternative methods of rapid and accurate diagnosis of patients with COVID-19.

Based on initial chest CT scans and associated clinical information (including epidemiological history, leukocyte counts, symptomatology, patient age and sex),  the authors~\cite{mei2020artificial} designed a deep learning based model to identify COVID-19 infection that could rapidly identify COVID-19 positive patients in the early stages. A deep convolutional neural network (CNN) was first developed to learn the imaging characteristics of COVID-19 patients on the initial CT scan. The support vector machine (SVM), random forest model, and multilayer perceptron (MLP) classifiers were then used to classify COVID-19 patients based on clinical information, while MLP showed the best performance on the tuning set and only MLP performance would be reported hereafter. Finally, a neural network model combined with radiology data and clinical information was generated to predict COVID-19 infection status. The generated models were evaluated on the test set, and their performance was compared to one fellowship-trained thoracic radiologist and one thoracic radiology fellow (Table~\ref{table:models}). Two-sided P values were calculated by comparing the sensitivity, specificity, and area under the curve (AUC) between each of the two models. The CIs of AUC were calculated with DeLong methods~\cite{delong1988comparing} for evaluation. The sensitivity and specificity comparisons were calculated via the exact Clopper-Pearson method~\cite{agresti1998approximate} to compute the 95\% CI shown in parentheses and exact McNemar’s test	~\cite{mcnemar1962psychological} for P value.

\begin{table}[ht]
	\centering
	\caption{Models in \cite{mei2020artificial}}
	\scalebox{0.95}{
		
		\begin{tabular}{ | c | c | c | c | c |}				
			\hline
			 &  \makecell{Positive \\ but \\ normal CT} & AUC, P & Sensitivity, P  &  Specificity, P  \\ \hline	

\makecell{Senior \\ Thoracic \\ Radiologist} & 0 / 25 & \makecell{0.84 \\ (80.0, 88.4), \\ N/A }& \makecell{0.746 \\ (66.4, 81.7), \\ P $=$ 0.0501} & \makecell{93.8 \\ (88.5, 97.1), \\ P $=$ 0.005} \\ \hline 

\makecell{Thoracic \\ radiology \\ fellow} & 0 / 25 & \makecell{0.73 \\ (68.3, 78.0), \\ NA } & \makecell{0.560 \\ (47.1, 64.5), \\ P $=$ 0.0004} & \makecell{90.3 \\ (84.3, 94.6), \\ P $=$ 0.090 }\\ \hline

CNN Model & 13 / 25 &  \makecell{0.86 \\ (0.821, 90.7), \\ P $=$ 0.00146} & \makecell{0.836 \\ (76.2, 89.4), \\ P $=$ 1.00} & \makecell{75.9 \\ (68.1, 82.6), \\ P$=$0.031} \\ \hline

MLP Model & 16 / 25 & \makecell{0.80 \\ (0.746, 84.9), \\ P $=$ 0.0004} & \makecell{0.806 \\ (72.9, 86.9), \\ P $=$0.442} & \makecell{0.683 \\ (60.0, 75.8), \\ P $=$ 0.0004} \\ \hline

Joint Model & 17 / 25 & \makecell{0.92 \\ (88.7 94.8), \\ NA} & \makecell{84.3 \\ (0.771, 90.0), \\ NA} & \makecell{82.8 \\ (0.756, 88.5), \\ NA} \\ \hline

		\end{tabular}	
	}
	
	\label{table:models}
\end{table}

The proposed joint AI algorithm~\cite{mei2020artificial} combined with both clinical data and CT imaging performed well in sensitivity (84.3\%) and specificity (82.8\%), and achieved an AUC of 0.92. It can be hypothesized that AI systems will help to rapidly diagnose COVID-19 infected patients when chest CT scans and associated clinical history are available, and therefore help in training the health system and combating the COVID-19 pandemic. 

Based on the images produced by X-rays and CT scans, researchers attempted to design COVID-19 specific deep neural networks to increase the accuracy of the diagnosis~\cite{wang2020covid, huang2020serial, wang2020deep, apostolopoulos2020covid}. 
Due to very limited data sets, authors of~\cite{apostolopoulos2020covid} used 
transfer learning to train the deep CNNs. Firstly, they applied transfer learning on different CNNs models, such as VGG-19~\cite{simonyan2014very}, MobileNets V2~\cite{howard2017mobilenets}, Inception V4~\cite{szegedy2017inception} and Xception~\cite{chollet2017xception}. Then, the best two models on accuracy, MobileNet v2, and VGG-19 were selected for 
COVID-19 classification, which involves 224 images with positive Covid-19, 700 images with confirmed common bacterial pneumonia, and 504 images without diseases.

\begin{table}[ht]
	\centering
	\caption{Models in \cite{apostolopoulos2020covid}}
	\scalebox{0.95}{
		
		\begin{tabular}{ | c | c | c | c | c |}				
			\hline
			Model &  \makecell{Predicted \\ labels} & \makecell{Actual \\ COVID-19}  &  \makecell{Actual \\ Pneumonia} & \makecell{Actual \\ Normal}  \\ \hline	
			
			MobileNet v2 & \makecell{ Covid-19 \\ Pneumonia \\ Normal} & \makecell{ 222 \\ 2 \\ 0} & \makecell{ 8 \\ 495 \\ 1} & \makecell{ 27 \\ 27 \\ 646} \\ \hline 
			
			VGG-19 & \makecell{ Covid-19 \\ Pneumonia \\ Normal} &  \makecell{ 222 \\ 3 \\ 13} & \makecell{ 8 \\ 460 \\ 36} & \makecell{ 7 \\ 26 \\ 667} \\ \hline 
		 
		\end{tabular}	
	}	
	\label{table:workload}
\end{table}

Furthermore, COVID-Net~\cite{wang2020covid} makes predictions using a design to fully understand the critical factors associated with positive cases, which
helps clinicians to improve screening and in the meantime, audit COVID-Net in a responsible and transparent manner to ensure that only relevant information from the CXR images is leveraged in the decision making. 

While many efforts have been made to utilize artificial intelligence assisted analysis in combating COVID-19, the biggest challenge in the field is the shortage of data sets. We summarize the existing data sets that are publicly available below.

\begin{itemize}
\item COVIDx~\cite{COVIDx}: It is a combined data set from five different sources that contain Chest radiography images of 7966 normal, 5451 Pneumonia, 258 COVID-19 patients. 

\item Italian radiological cases~\cite{irc}: It contains 115 COVID-19 patients with detailed symptomatography and images at different stages.

\item BIMCV-COVID19+~\cite{BIMCV-COVID-19}: It is a large data set with chest X-ray images and CT imaging of COVID-19 patients along with their radiographic findings, pathologies, polymerase chain reaction, immunoglobulin G and immunoglobulin M (IgM) diagnostic antibody tests, and radiographic reports. Currently, it includes 1380 CX, 885 DX, and 163 CT studies.

\end{itemize}

\section{Policy Responses and Effectiveness}
\label{policy}

As the COVID-19 pandemic continues to spread, governments and international organizations 
are implementing various policies, which aim to deliver systematic, effective, and coordinated responses
to flatten the curve, save lives and restart the economy. The following items summarize basic policy responses 
that were widely implemented by the governments globally.

\begin{itemize}

\item Social distancing: Keep space (e.g. 6 feet) between yourself and other people outside of your home.
It means the reduced capacity for indoor businesses and activities, such as restaurants and schools.

\item School closures: In most schools, it is impossible to maintain social distancing in the classroom. Most of the schools were closed in response to the pandemic and transitioned to online lecturing.

\item Travel restrictions: Stop non-essential travels, travel bans on specific counties or regions, border closures, e.g. US-Canada and US-Mexico closed on March 18.

\item Face covering and mask requirement: Cloth face coverings are required when not working alone and when interacting with the public, masks should be worn.

\item Stay in the home: Except essential workers, all should remain at home and away from other people unless it is absolutely necessary to go out (e.g. grocery shopping and doctor visit). Note California has a similar policy named shelter in place.

\item Phased reopening: Based on the government evaluations, reopening the economy following a phased structure such that
each phase remains around two weeks for further evaluation.

\end{itemize}

\begin{figure*}[!t]

\centering
         \includegraphics[width=\linewidth]{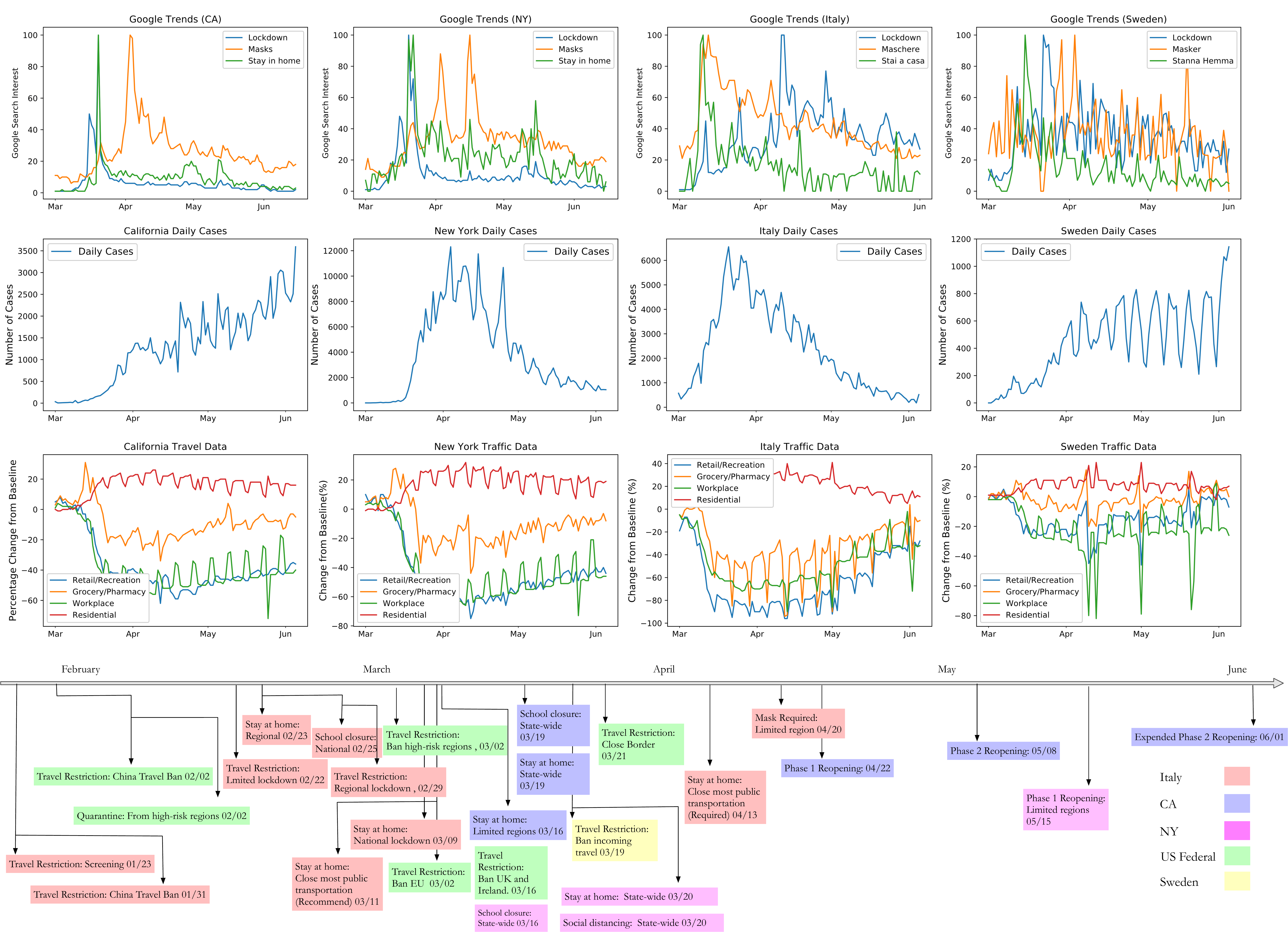}           
\caption{Comparison of Google Trends, Community Traffic and Daily Infections of New York, California, Italy and Sweden along with the timeline of the policies}  
\label{policy-fig}    
\end{figure*}

Fig.~\ref{policy-fig} plots the timeline of key police interventions from government of New York (NY) State, California (CA) State, Italy, Sweden and United States Federal along with the commonly used analytical data sources from Google Coronavirus Search Trends~\cite{searchtrend}, Daily Infection Curves~\cite{jhu} and Google Community Mobility Report~\cite{cmr}. The Google Search Interest demonstrates the degree of the propaganda that each region involved, where the values of interests stay at high level in Italy starting from early March,
but in NY and CA the interest started jittering in mid-March. 
When various policies, such as different levels of stay-at-home order, implemented in these regions, the community mobility decreased quickly in NY, CA and Italy for workplaces, retail/recreation, which were not recommended or prohibited under the order.
The degrees of decreases can reflect the level of restrictions, for example, in Sweden, there were roughly 35\% reduce in workplaces, however, the value in Italy was 75\%. This is because Sweden implemented a partial stay-at-home policy, which only recommend vulnerable people (e.g. seniors) to stay at home.

\subsection{Social Distancing Policies}

The mobility data is used to gauge the effectiveness of social distancing and stay at home orders, as well as how well they were followed. The authors in~\cite{abouk2020immediate} ranked different intervention policies based on their effectiveness, by using a difference-in-differences methodology, location-based mobility, and daily state-level data COVID-19 tests and confirmed cases. The mobility data collected by the Google Community Mobility Reports~\cite{cmr} was split into county-level data and state-level data. Since most of the intervention policies were implemented state-wide, the authors~\cite{abouk2020immediate} performed an analysis based on the state-level data and had already collected movements for 50 states and the District of Columbia for 29 days with 1479 observations. The web-scraping daily temperature data was captured for the top 5 biggest cities in each state from Weather Underground~\cite{wu} (commercial weather service with real-time weather information). The data for the daily state-level numbers of tests and positive cases were collected from the COVID Tracking Project website from March 9 to April 20, 2020, and the author~\cite{abouk2020immediate} had data on all 50 states and the District of Columbia for 43 days, providing 2193 observations. 

The linear regression model and a difference-in-differences methodology were used to evaluate the effect of the COVID-19 policies by authors~\cite{abouk2020immediate}. A binary variable was defined for each policy, set to one if a given state adopts that policy after a certain day during the sample period, and otherwise to zero. The regression equation is: 

\begin{equation}
Y_{s t}=\alpha+X_{st} \beta+\omega * \operatorname{temp}_{st}+\delta_{s}+\tau_{t}+\varepsilon_{s t}
\end{equation}
, where $Y$ is the changes in visiting various places; X represents the matrix for COVID-19 policies, $temp$ indicates state-level mean daily temperature, $\delta$ and  $\tau$ are sets of state and day-of the month fixed effects. $\alpha$ and $\beta$ are the fitting coefficients.

When estimating the effect of COVID-19 policies on the number of confirmed cases, it studied the Poisson regression model:

\begin{equation}
\begin{split}
Pos_{s t}=\exp \left(\alpha+\sum_{\tau<=-7}^{\nabla>=15} \beta_{\tau} X_{\tau, s t} + \omega \times \operatorname{tem} p_{s t}
\right.
\\
\left. + \lambda \times \log (\text {tests}+1)+\delta_{s}+\tau_{t} \vphantom{\int_1^2} \right)
\end{split}
\end{equation}
,where $Pos_{st}$ is the state-level daily number of confirmed cases. Since the confirmed cases in each state heavily depended on the number of conducted COVID-19 tests, the log-transformed version of the test number variable was used to interpret the estimated coefficient as elasticity. The results demonstrated that statewide stay-at-home orders significantly increased the measure associated with presence at home by about six fold  (relative to states without policy). Though the policies such as non-essential business closures and restaurant and bar limits have positive and statistically significant impact on presence at home, their effect sizes were about half of what observed for stay-at-home orders. Meanwhile, there existed a steady decline in the number of daily confirmed COVID-19 cases after 10-15 days after such policies were implemented.

Similarly, a research group in University of Wisconsin-Madison~\cite{gao2020mobile} used two social distancing metrics.
\begin{itemize}
\item The median of individual maximum travel.
\item The home dwell time.
\end{itemize}
The data are derived from large-scale mobile phone location data provided by Descartes Lab~\cite{descarteslabs} and SafeGraph~\cite{safegraph}.
The metrics are used to evaluate the effectiveness of series of stay-at-home policing on decelerating the spread of the COVID-19 epidemic by mathematical curve fitting models and mechanistic epidemic prediction models. Their results~\cite{gao2020mobile} confirmed that state implemented stay-at-home orders increased the amount of time spent at home and the increasing stay-at-home dwell time would help to decrease the amount of daily cases of COVID-19.  In conclusion, both studies~\cite{gao2020mobile, abouk2020immediate} confirmed that the amount of positive daily COVID-19 cases decreased as more stay-at-home policies were implemented, with the stay-at-home orders being the most effective and bar and restaurant closings being the least effective.

By using metro traffic data to compare epidemics in two major cities with the largest number of COVID-19 reported cases (Daegu and Seoul), the authors in~\cite{park2020potential} described potential roles of social distancing in mitigating the spread of COVID-19 in South Korea. The authors collected daily numbers of reported cases data in two geographic regions from the Korea Centers for Disease Control and Prevention (KCDC) between January 20 to March 16, 2020, and the daily metro traffic in two cities between 2017 to 2020 was obtained from data.go.kr and data.seoul.go.kr.

The time-dependent reproduction number $R_t$, which represents the average number of secondary cases caused by an average individual, given conditions at time t, was estimated using the following equation with a 14-day sliding window: 
\begin{equation}
R_{t}=\frac{I_{t}}{\sum_{k=1}^{14} I_{t-k} w_{k}}
\end{equation}
,where $I_t$ is the reconstructed incidence time series, for example, the number of infected cases on day $t$, and $w_k$ represents the generation-interval distribution randomly drawn from a prior distribution.

After comparing the reconstructed incidence and estimates of $R_t$ in Daegu and Seoul, the results showed that the estimates of $R_t$ gradually decreased and eventually dropped below 1 about one week after the reporting of the first case, while the metro traffic volume also decreased simultaneously. The clear, positive correlations between the normalized traffic and the median estimates of Rt were found in both Daegu ($r$ = 0.90; 95\% CI: 0.79-0.95) and Seoul ($r$ = 076; 95\% CI: 0.59-0.87), which indicated that staying away from the metro and traveling less had a positive correlation with preventing spreading the virus.

\subsection{Travel Restriction, School Closure and Large-scale Lockdown}

To reduce the spread of COVID-19 pandemic in China, restrictions on mobility (hereafter called cordon sanitaire) were imposed on Wuhan City, Hunbei province on January 23, 2020~\cite{kraemer2020effect}. To elucidate the role of case importation in transmission in cities across China, the authors collected real-time mobility data from Baidu Inc., together with epidemiological data from each province, and detailed case data with reported travel history. These data would help to ascertain the impact of control measures. Three different COVID-19 "Generalized" Linear Models, GLM, were built to evaluate hypotheses regarding the effect of mobility and testing on COVID-19 dynamics; model 1 and model 2 were a Poisson GLM and a negative binomial GLM to estimate daily cases counts, where model 3 used a log-linear regression to estimate daily cumulative cases. 

The findings in~\cite{kraemer2020effect} confirmed that the travel restrictions were particularly helpful in the early stage of an outbreak when it was more confined but became less effective as the outbreak became more widespread. The real-time human mobility data from Baidu Inc. presented an expected decline of importation after the establishment of the cordon sanitaire. Since the travel bans prevented traveling into and out of Wuhan around the time of the Lunar New Year celebration, the bans may help to reduce further dissemination of COVID-19 from Wuhan. Except for Hubei, the study also estimated COVID-19 growth rates in all other provinces and found that all other provinces experienced faster growth rates before travel restrictions and substantial control measures were implemented. After the control measures were implemented, growth rates became negative.

The authors in~\cite{anzai2020assessing} used the example of Japan, the country in Asia that received the largest number of visitors from China, to quantify the impact of the drastic reduction in travel volume on the COVID-19 transmission dynamic outside China, and to estimate reduction in COVID-19 infections and  the chance of an outbreak outside China as a result of such travel policies. The epidemiological datasets of confirmed COVID-19 cases outside China were collected from government and news websites as of February 6, 2020. The author [35] quantified the impact on the reduced number of exported cases, the reduced probability of a major epidemic overseas, and the time delay to a major epidemic gained from the reduction in travel volume. The author [35] assumed the epidemic start data was set on December 1, 2019 (Day 0), and then Wuhan was put in lock-down from Day 53 (January 23, 2020). Since the mean incubation period of COVID-19 was nearly 5 days, thus, the impact of reduced travel volumes would start to be interpretable from Day 58. The counterfactual model was used to estimate the reduced volume of exported cases, and Poisson regression was used to fit the following model through Day 57 with following equation:
\begin{equation}
E(c(t))=c_{o} \exp (r t)
\end{equation}
where $c(t)$ was the incidence of exported cases on Day $t$, $c_o$ was the initial value at $t = 0$ and $r$ presents the exponential growth rate of exported cases outside China. The reduced travel volume of exported cases by Day 67 was calculated by,

\begin{equation}
V=\sum_{t=58}^{67}(h(t)-E(c(t)))
\end{equation}
, where $h(t)$ showed the observed number of cases on day $t$.

According to the calculations and predicted curve, the expected number of confirmed COVID-19 cases between Day 58 (January 28, 2020) and Day 67 would be 321 (with 95\% confidence interval: 181, 544), and a total of 95 cases were diagnosed in the empirical observation. Based on the results, the authors estimated that 226 cases (95\% CI: 86, 449) were prevented from being exported across the world as a result of the Wuhan lockdown. Furthermore, the researchers considered the probability of a major pandemic and the possible delay, specifically focusing on Japan. Without travel restrictions, researchers found that the probability of a major pandemic would be more than 90\%, while it would be "broadly ranged from 56\% to 98\%" with restrictions. When mobility is limited, the delay (in days) in time to pandemic is decreased. Furthermore, this paper considers the reduction in COVID-19 spread through contact tracing, where risk reduction reached 37\% when 50\% of those infected were traced. In Japan, researchers found that the probability of a major epidemic was estimated to be reduced by 7\%-20\% and a 2-day delay was gained in the estimated time to a major epidemic.

\subsection{Long-term Impact of COVID-19 policies}
\begin{figure}[!t]
\centering
 \includegraphics[width=\linewidth]{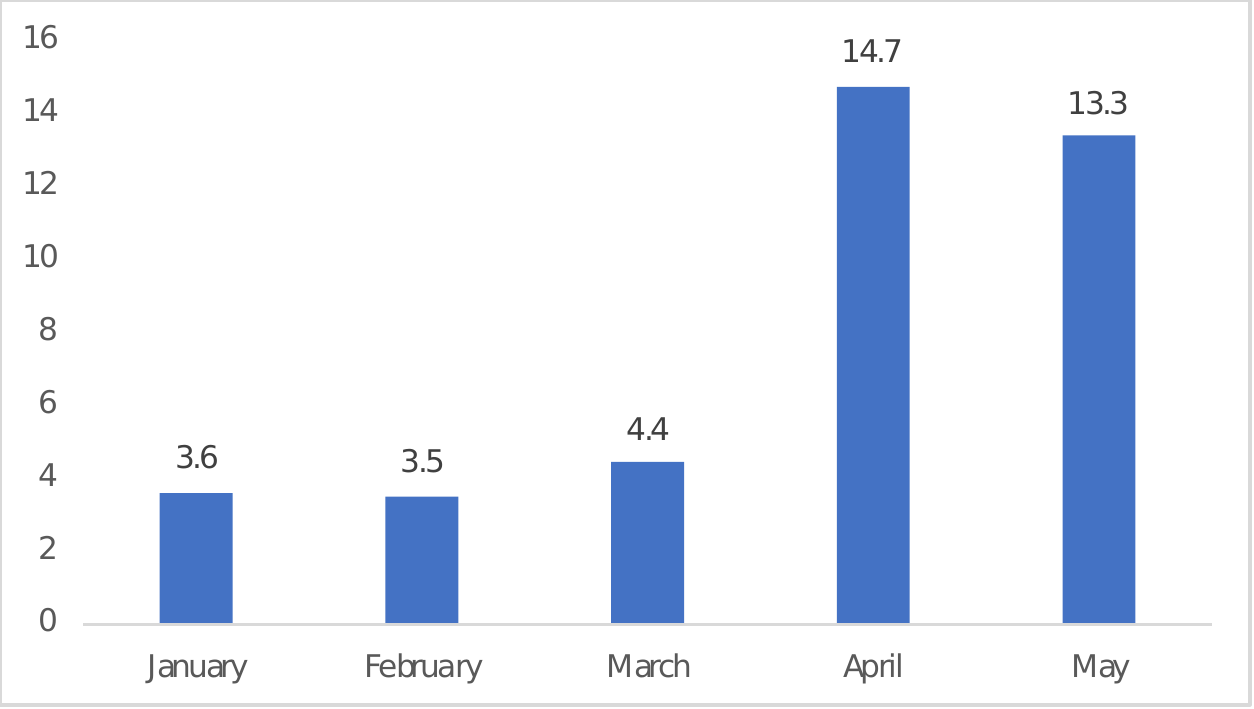}           
\caption{Unemployment Rates in United States 2020}  
\label{unemployment}    
\end{figure}

With nearly every state in the United States placing stay-at-home order and shutting down schools for the rest of the 2019 - 2020 academic year due to the COVID-19 pandemic. The long-term effect of these policy responses attracts many researchers. 
The Fig.~\ref{unemployment} plots the unemployment rate in United States that can reflect the immediate economic impacts of the policies.

During the pandemic, many employees are unable to travel to work. The authors in~\cite{dingel2020many} investigate
on how many jobs can be done at home based the data collected by the Occupational Information Network (O*NET) surveys that covered "work context" and "generalized work activities".  
They also took this data and merged it with data from the Bureau of Labor Statistics to show the prevalence of each job listed in the United States.  
The results from this paper showed that about 37\% of United States jobs can be done from home, which account for 46\% of all wages.  

There are more than 55 million students are out of school without an explicit expectation of school reopening.  Yet, education leaders have little information on how the education system has been impacted by school closures. How to model the potential impact of COVID-19 school closure based on the existing data becomes a critical topic in the field of educational policy.

To project the learning loss caused by school closure, the authors in~\cite{basilaia2020transition} assumed that the learning loss due to COVID-19 can be deemed as an extended learning loss due to summer break. They used the data from the past MAP Growth assessment takers (Grades 3-8 students taking exams in 2017-18 and 2018-19 school year) to estimate the average summer loss. By using the "typical" school year growth rates and summer loss as a reference, they built regression models to project the learning loss due to COVID-19. Under these models, it is estimated that the students would obtain approximately 63 - 68\% of the learning gains in reading relative to a typical school year and with 37-50\% of the learning gains in math if they were able to return schools in 2020 Fall. The worry in learning loss also exists in worldwide. The article~\cite{rti} analyzed 27 datasets from low- and middle-income countries to estimate year-on-year growth in student reading achievement under normal conditions. They assumed that learning loss can be estimated as a constant relative to the percent of schooling lost. (e.g. grade 3 students were consistently reading about 20 words per minute faster than their grade 2 counterparts at given percentiles in consecutive grades.) While a 30\% learning loss (i.e. the equivalent of an approximately 3-month school closure) would yield a 5.9 correct word per minute loss for mid-percentile grade 3 students (30\% of the 19.8 correct word per minute expected gain).

Predicting the long-term effect of COVID-19 is still in a very early stage, related communities are encouraged to collect data from various sources.

\section{Contact Tracing}
\label{contact}

In public health, contact tracing is the process of identification of persons who may have come into contact with an infected person and subsequent collection of further information about these contacts. 
In practice, however, it is a challenging task to record the close contact (e.g. 6 feet) through daily routine intersects.

The researchers are rapidly coalescing around applications for proximity tracing. Different technologies are been utilize in this field. For example, the Bluetooth signal strength can be used to determine whether two smartphones were close enough together for their users to transmit the virus. 

The two dominant mobile operating systems owners, Apple and Google published Exposure Notification (a.k.a Privacy-Preserving Contact Tracing Project) in late April. It is a system that contains public available specifications developed by Apple and Google. Exposure Notification utilizes Bluetooth Low Energy technology and privacy-preserving cryptography to decide whether a specific user may have recently been within the proximity of someone that had been infected with COVID-19.
Due to security, privacy and political concerns, however, some governments (e.g. Norway, France and United Kingdom) tend to develop their own version the application. We summarize the mainstream contact tracing application below.

\begin{itemize}

\item Singapore, TraceTogether~\cite{tracetogether}: it uses Bluetooth to approximate your distance to other phones running the same app and stores data for up to 25 days. It does not collect GPS locations or data about users' WiFi or mobile network.

\item China, Chinese health code system~\cite{chinacode}: it is built inside two hugely popular applications WeChat and Alipay in China, to provide
a health survey and location based colored health code. The mobile network association is collected at backend to track users location.
 
 \item Austria, StoppCorona~\cite{stopp}: it is an open-source project for bluetooth based contact tracing. It claims to use a  decentralized approaches for the tracing.

\item Hong Kong, StayHomeSafe~\cite{stayhomesafe}: the application together with a wristbands, which is given to  all arrivals at the airport, is 
used to strictly enforce 14-day quarantine. The users need to scan an unique QR code to pair the wristband with the app. Once home, they are to walk around the apartment to calibrate the wristband.

\item South Korea, Corona 100~\cite{covid100}: it utilizes government data, alerts users when they come within 100 meters of a location visited by an infected person. The GPS data is used to keep tracking the users' location.

\item France, STOPCovid~\cite{STOPCovid}: it relies on Bluetooth Low Energy to build record the users nearby. If a user test positive of COVID-19, he/she would get a QR code from the doctor and the user can choose to open the app and enter that code to notice the people that he/she interacted with over the past two weeks.

\item Japan, COCOA~\cite{cocoa}: it is developed by a group of engineers at Microsoft and utilizes Exposure Notification platform. It records encrypted data flagging phones that have been within one meter for more than 15 minutes; when one person reports the fact that they have tested positive for COVID-19, those other users will be notified.

\item  India, Aarogya Setu~\cite{aarogyasetu}: based on the both Bluetooth and GPS technologies, it lets users know if they have been near a person with Covid-19 by scanning a database of known cases of infection. The gathered the data stores on the servers and shared with the government.

\item Italy, Immuni~\cite{immuni}: it follows the standards of Exposure Notification, which uses Bluetooth to swap codes between mobile devices.  

\item Norway, Smittestopp~\cite{smittestopp}, it utilizes both Bluetooth and GPS singles to estimate user proximity as a means of calculating exposure risk to COVID-19. In addition, it is a centralized application architecture, which means the data is uploaded to a central server controlled by the health authority, instead of being stored locally on devices.

\item United Kingdom, NHS COVID-19 App~\cite{nhs}: it leverages a centralized design that  uses Bluetooth to trace the users and stores data on NHS's servers. (UK is in the transition to move the application under Exposure Notification specifications.)

\end{itemize}

App-based contact tracing is necessary and useful to control the COVID-19 pandemic since not enough to quarantine people only after symptoms onset. To reduce infections, when a person is confirmed with COVID-19 infection, one should act quickly to find all people this person was in close proximity with. Only a digital, largely automatic solution would help to conduct such fast contact tracing. However, how to effectively evaluate these applications are still under investigation.

The authors in~\cite{ferretti2020quantifying} used published parameters for the incubation time distribution (5.2 days) and the epidemic doubling time (5.0 days) from the early epidemic data in China to develop a mathematical model for COVID-19 infectiousness to analyze the contribution of different transmission routes. The model estimated the basic reproductive number $R_0$ equaled two in the early stages of the epidemic in China, while the contributions to $R_0$ included four parts 1) 46\% from presymptomatic individuals (who had not shown symptoms yet), 2) 38\% from symptomatic individuals, 3) 10\% from asymptomatic individuals (who never show symptoms), and 4) 6\% from environmentally mediated transmission via contamination. The general mathematical model of COVID-19 infectiousness was determined to illustrate the infectiousness varies as a function of time since infection, $\tau$ , for a representative cohort of infected individuals~\cite{ferretti2020quantifying}. The equation is presented as: 
\begin{equation}
\begin{split}
\beta(\tau)=P_{a} x_{a} \beta_{s}(\tau)+\left(1-P_{a}\right)[1-s(\tau)] \beta_{s}(\tau) \\ +\left(1-P_{a}\right) s(\tau) \beta_{s}(\tau)+\int_{l=0}^{\tau} \beta_{s}(\tau-1) E(l) d l
\end{split}
\end{equation}
,where $P_{a}$ is proportion asymptomatic, $x_{a}$ represents relative infectiousness of asymptomatics, $\beta_{s}(\tau)$ describes the infectiousness of an individual currently either symptomatic or presymptomatic, at age of infection $\tau,$ and $E(l)$ presents environmental infectiousness, which indicates the rate at which a contaminated environment infects new people after a time lag $l$.

In order to estimate the requirements for successful contact tracing, the authors~\cite{fraser2004factors} determined the combination of two key parameters needed to reduce $R_0$ to less than 1: 1) the symptomatic individuals should be isolated and 2) the contacts of symptomatic cases should be traced and quarantined. Based on published analytical mathematical framework [2], the authors in [1] quantified the whether the COVID-19 epidemic was expected to be controlled or not by these two interventions. The results indicated that if used by a sufficiently high proportion of the population, immediate notification through a contact-tracing mobile phone app could be sufficient to stop the epidemic. Practical and logistical factors including uptake, coverage, R0in a given population would be used to determine whether an app is sufficient to control epidemic spread, or whether additional measures are required to reduce R0. The performance of the app can be explored at~\cite{routes}. 

The authors in~\cite{kretzschmar2020isolation} estimated the conditions that isolation and contact tracing in settings with various levels of social distancing would be able to contain or slow down COVID-19 epidemic. A stochastic transmission model in~\cite{kretzschmar2004ring} was used to calculate the numbers of latently infected persons, infectious persons, and persons who have been diagnosed and isolated in time steps of one day. The author used the model to distinguish between household contacts (close contacts) and non-household contacts, and found that only if the majority of cases were ascertained, then isolation and contacting tracing would be an effective methods to slow down epidemics. Meanwhile, social distancing would reduce the effective reproduction number to below one when non-household contacts were reduced by around 90\%. Finally, the combination of social distancing with isolation and contacting tracing have synergistic effects that would increase the prospect of containment.

While many countries utilize Bluetooth-based technologies~\cite{mao2016mobile, mao2014pasa} to help slow the spread, digital contact tracing comes with serious privacy concerns because many proposed apps rely on geo-location tracking and some of them store user data on central servers if people are to be identified and tracked. And due to the lack of privacy regulations by the government, users have to depend on the good will of technology companies to avoid violating their privacy~\cite{bradford2020covid, zastrow2020coronavirus}.



\section{Conclusion}
\label{con}

As the COVID-19 pandemic continues, many academic papers have been published to 
help with combating it. In this paper, we conduct a literature review from the perspective of
data-driven analytics. We investigate the latest solutions for epidemic prediction models, clinical diagnosis, policy
effectiveness and contact tracing. Additionally, we study models with latest data to evaluate how
good they perform since their publication date and collect data
sources for analytical researches.

\bibliographystyle{IEEEtran}
\bibliography{routing}

\end{document}